\documentclass[11pt,dvips]{article}

\usepackage{jheppub}

\usepackage{slashbox,latexsym,amssymb,amsfonts,amsmath,slashed}
\usepackage{graphicx}
\newcommand{\f}[2]{\frac{#1}{#2}}
\newcommand{\la}{\langle}

\newcommand{\ra}{\rangle}

\newcommand{\slaD}{{\slashed D}}
\newcommand{\de}{\partial}
\newcommand{\nyp}[3]{{\bf #1} (#3) #2}
\newcommand{\JHEP}{Jour.\ High Energy Phys.\ }

\title{An Ising-Anderson model of localisation in high-temperature
  QCD}

\author[a]{Matteo Giordano}
\author[a]{\!\!, Tam\'as G.\ Kov\'acs}%
\author[b]{\!\!, and Ferenc Pittler}%

\affiliation[a]{
Institute for Nuclear Research of the Hungarian Academy of Sciences, \\
Bem t\'er 18/c, H-4026 Debrecen, Hungary 
}%
\affiliation[b]{
  MTA-ELTE Lattice Gauge Theory Research Group,\\ P\'azm\'any
  P. s\'et\'any 1/A, H-1117 Budapest, Hungary}

\emailAdd{giordano@atomki.mta.hu}
\emailAdd{kgt@atomki.mta.hu} 
\emailAdd{pittler@bodri.elte.hu}

\abstract{
  We discuss a possible mechanism leading to localisation of the
  low-lying Dirac eigenmodes in high-temperature lattice QCD, based on
  the spatial fluctuations of the local Polyakov lines in the
  partially ordered configurations above $T_c$. This mechanism
  provides a qualitative explanation of the dependence of localisation 
  on the temperature and on the lattice spacing, and also of the phase
  diagram of QCD with an imaginary chemical potential. To test the
  viability of this mechanism we propose a three-dimensional
  effective, Anderson-like model, mimicking the effect of the Polyakov
  lines on the quarks. The diagonal, on-site disorder is governed by a
  three-dimensional Ising-like spin model with continuous spins. Our
  numerical results show that localised modes are indeed present in
  the ordered phase of the Ising model, thus supporting the proposed
  mechanism for localisation in QCD.  
}


\begin{document}

\maketitle

\section{Introduction}

The spectrum of the lattice Dirac operator plays a prominent role in
current attempts to improve our understanding of the spontaneous
breaking of chiral symmetry in QCD. The key relation in this context
is the celebrated Banks--Casher formula~\cite{BC}, which clarifies the
relevance of the low-lying Dirac eigenmodes for the development of a
non-vanishing chiral condensate in the chiral limit. The Banks--Casher
relation also suggests that the (pseudo)critical behaviour of the
theory at the chiral transition/cross-over, even away from the chiral
limit, will be mostly determined by the behaviour of these modes.

It has long been known that in QCD, below the pseudocritical
temperature, $T_c$, the low-lying eigenmodes of the Dirac operator are
delocalised, i.e., they extend throughout the whole four-dimensional
volume of the system~\cite{VWrev} (see also Ref.~\cite{deF}). 
In recent years it has been observed that above $T_c$
this is no longer true: the lowest-lying modes are localised on the
scale of the inverse temperature~\cite{KP2,feri}. Delocalised
modes are still present, but only above a (temperature-dependent)
critical point $\lambda_c=\lambda_c(T)$ in the spectrum. 
A similar change in the properties of the low-lying modes is found
also in other, QCD-like theories, e.g., with $SU(2)$ gauge group, and/or in
the quenched limit~\cite{GGO,GGO2,KGT,KP}.  

As the temperature is decreased, $\lambda_c$ moves towards the origin,
and vanishes at a temperature compatible with $T_c$, the cross-over
temperature determined from thermodynamic
observables~\cite{Aoki:2005vt,Borsanyi:2010cj}. This is certainly not
a coincidence. First of all, in QCD~\cite{GGO2,KP2} and in all the
QCD-like theories where localised modes have been observed (pure-gauge
$SU(2)$~\cite{KGT,KP} and $SU(3)$~\cite{GGO2} theories), they are
present only in the  ``chirally-symmetric phase'', independently of
the specific lattice discretisation of the Dirac operator
(staggered~\cite{GGO2,KP}, overlap~\cite{KGT}, and recently also
M\"obius domain wall~\cite{Cossu:2014aua}). Moreover, further evidence
of a very close relationship between the appearence of localised modes
and the chiral transition has been found recently~\cite{GKKP} in a toy
model for QCD, namely $N_T=4$ unimproved staggered
fermions~\cite{unimproved}, which displays a genuine first-order
chiral transition accompanied by the appearence of localised modes at
the low end of the spectrum. However, the precise nature of this
relationship has not been fully understood yet. 

At fixed temperature, the transition in the spectrum from localised to
delocalised modes has been shown to be a true, second-order
Anderson-type transition~\cite{crit}. Such transitions have been most  
extensively studied before using the Anderson
model~\cite{Anderson58,LR,EM}, which is a model for electrons in a
crystal with disorder. The Hamiltonian of the Anderson model consists
of the usual tight-binding Hamiltonian plus an on-site (diagonal)
random potential mimicking the presence of impurities in the
crystal. In three dimensions, when the diagonal disorder is switched
on, localised modes appear at the band edges, beyond critical energies
called ``mobility edges''. As the width of the distribution of the
diagonal disorder is increased, i.e., as the system is made more
disordered, the mobility edges move towards the band center, and
beyond a critical value for the disorder all the modes become
localised.   

The model described above is the so-called {\it orthogonal} Anderson
model. A variant of this model is the {\it unitary} Anderson model, in
which the hopping terms are modified with the introduction of random
phases to account for the presence of a magnetic field.
The specifiers ``orthogonal'' and ``unitary'' refer to the symmetry
class of the model in the language of random matrix theory
(RMT)~\cite{Mehta}.  
In this classification, QCD belongs to the unitary class. 
RMT makes universal predictions for certain statistical properties of
the spectrum of a random Hamiltonian with i.i.d.\ entries, depending
on its symmetry class. These predictions correctly describe also the
spectral region corresponding to the delocalised modes of the Anderson
Hamiltonian, despite this being a sparse matrix. However, this can be
understood by noticing that delocalised eigenmodes are easily mixed by
fluctuations in the disorder. On the other hand, localised modes are
sensitive only to fluctuations taking place within their support, and
thus they hardly mix. As a consequence, the corresponding eigenvalues
are expected to fluctuate independently, thus obeying Poisson
statistics. This behaviour is actually observed in the localised part
of the spectrum of the Anderson model. Therefore, a change in the
spectral statistics takes place in correspondence to the
localisation/delocalisation transition.  

The staggered Dirac operator in lattice QCD, being anti-Hermitian,
admits a straightforward interpretation as a random Hamiltonian. As in
the Anderson model, the statistical properties of the spectrum change
from Poisson-type in the spectral region of localised modes, to the
appropriate RMT-type in the region of delocalised modes. Besides this,
however, at first sight one can find very little in common with the
three-dimensional Anderson model. For one thing, the dimensionalities
apparently do not match. Moreover, the Dirac operator has only
off-diagonal non vanishing terms, i.e., the fluctuating hopping terms
representing the gauge fields. In this case the location of the
mobility edge was found to be controlled by the physical temperature.  
Finally, unlike in the Anderson model, the disorder in the gauge links 
is correlated. It is therefore surprising that the critical exponent
of the correlation length has been found to be compatible with that of
the unitary Anderson model in three dimensions~\cite{nu_unitary},
suggesting that the two models belong to the same universality
class~\cite{crit}. The unitary class is indeed the symmetry class
appropriate for QCD, and correlations in the disorder should not
affect the critical behaviour since they are short-ranged, as the
disorder consists of gauge-field fluctuations. On the other hand, the
fact that a four-dimensional model with only off-diagonal disorder
behaves like a three-dimensional one with diagonal disorder calls for
an explanation.\footnote{Although the unitary Anderson model contains
  also off-diagonal disorder, this is known to be much less effective
  in producing localisation~\cite{offdiag,offdiag2}.}  

An important step towards clarifying these issues would be the
determination of the mechanism responsible for the localisation of the
low-lying modes. It was initially suggested~\cite{GGO} that instantons
could play  the role of localising ``defects'', but numerical
investigations by one of the authors and
collaborators~\cite{Bruckmann:2011cc} found a large mismatch between
the density of localised modes and the density of instantons, thus 
making this explanation unlikely. In the same paper, an alternative
proposal was made to explain localisation at high temperature, which
involves the behaviour of local fluctuations of Polyakov loops above
$T_c$. The basic idea is that, in the ordered phase, local
fluctuations of the Polyakov loop provide a localising ``trap'' for
eigenmodes. This is because the ``islands'' where the Polyakov loop
fluctuates away from the ordered value provide ``energetically''
favourable spatial regions for the quark eigenmodes to live
in. Numerical support to this proposal was also obtained, in the case
of the $SU(2)$ gauge group. 

The purpose of this paper is to refine this proposal, and extend it to
the $SU(3)$ case. This simple generalisation has nevertheless the
merit of clarifying some important aspects that went previously
unnoticed. To further test the viability of our explanation, we
construct a simple three-dimensional model which should display
localisation precisely through the mechanism proposed for QCD. Our
numerical results show that this is indeed the case, and that the
qualitative features of the eigenmodes and of the spectrum correspond
to those of the eigenmodes and of the spectrum of the Dirac
operator. In particular, our toy model displays a transition in the
spectrum from localised to delocalised modes, which is likely to
become a true phase transition in the thermodynamic limit.

The plan of the paper is the following. In Section \ref{sec:PL} we
discuss the mechanism for localisation in high-temperature QCD based
on fluctuations of Polyakov lines. In Section \ref{sec:IA} we
construct an effective model (``Ising-Anderson'' model) which should
produce localised states precisely through the mechanism proposed in  
Section \ref{sec:PL}. In Section \ref{sec:NR} we report on the
results of numerical simulations of the Ising-Anderson model. Finally,
in Section \ref{sec:concl} we state our conclusions and discuss
possible extensions of the present study.

\section{Polyakov lines and localisation}
\label{sec:PL}

In this Section we argue that high-temperature QCD contains an
effectively diagonal and three-dimensional disorder, which explains
why the critical behaviour at the localisation/delo\-cal\-i\-sa\-tion
transition in the spectrum corresponds to that of the
three-dimensional unitary Anderson model. 

An intuitive argument for the effectively reduced dimensionality
of high-temperature QCD is the following: as the size of the temporal
direction is smaller than the correlation length, the time slices are
strongly correlated and the system is effectively
three-dimensional. This implies that the Dirac eigenmodes should look
qualitatively the same on all time slices, in particular for what
concerns their localisation properties. 

This intuition is strengthened when we consider the role of the
antiperiodic boundary conditions in the temporal direction. To this
end, it is convenient to work in the temporal gauge, $U_4(t,\vec
x)=\mathbf{1}$ for $t=0,\ldots,N_T-2$ and $\forall\,\vec x$, in which
$U_4(N_T-1,\vec x)$ equals the local Polyakov line $P(\vec x) \equiv
\prod_{t=0}^{N_T-1} U_4(t,\vec x)$. A further time-independent gauge
transformation allows one to diagonalise each local Polyakov line: we
will refer to this as the diagonal temporal gauge. In any temporal
gauge, covariant time differences are replaced by ordinary differences
for all $t\neq N_T-1$. One can do the same also at $t=N_T-1$, if
at the same time one trades the antiperiodic boundary conditions for 
effective, $\vec x$-dependent boundary conditions, which involve the
local Polyakov line, 
\begin{equation}
  \label{eq:eff_bc}
\psi(N_T,{\vec x})=-P({\vec x})\psi(0,{\vec x})\,.  
\end{equation}
Since the time slices are strongly correlated, these effective,
$\vec x$-dependent boundary conditions will affect the behaviour at
the spatial point $\vec x$ for all times $t$. Furthermore, $P({\vec
  x})$ fluctuates in space (and obviously from one configuration to
another). From the point of view of a disordered system, QCD above
$T_c$ therefore contains effectively a diagonal (on-site),
three-dimensional source of disorder. 

To see how these effective boundary conditions affect the quark wave
functions, let us first discuss a simplified setting in which the
Dirac equation can be explicitly solved, generalising the argument of
Ref.~\cite{Bruckmann:2011cc} to $SU(3)$. We discuss here the na\"ive
lattice Dirac operator for simplicity, but the considerations of the
present Section clearly apply to the staggered Dirac operator as well,
since the two are essentially connected by a unitary transformation.
Consider configurations with constant temporal links $U_4(t,\vec x)=U$
and trivial spatial links $U_j(t,\vec x)=\mathbf{1}$, $j=1,2,3$. In
the diagonal temporal gauge, one has $P(\vec x)=P=U^{N_T}={\rm
  diag}(e^{i\phi_1},e^{i\phi_2},e^{i\phi_3})$ with
$\prod_{c=1}^3e^{i\phi_c}=1$, so the Dirac operator is diagonal in
colour, and the colour components of the quark eigenfunctions
decouple. The eigenfunctions of the na\"ive lattice Dirac operator 
have the following simple factorised form,
\begin{equation}
  \label{eq:eigen_simple}
  \begin{aligned}
 &   \slaD \psi^{(c k \vec p s\pm)} = \pm i\lambda^{(c k \vec p
      )}\psi^{(c k \vec p s\pm)}\,, &&&
&   \lambda^{(c k \vec p)} =  \sqrt{\sin^2\omega^{(c k)} + 
      \textstyle\sum_{j=1}^3\sin^2 p_j}\,, \\
 &   \psi_{\alpha j}^{(c k \vec p s\pm)}(t,\vec x) = 
    f^{(c k \vec p)}(t,\vec x)\,\chi_\alpha^{(c k \vec p s\pm)}\,\varphi_j^{(c)}\,,
  &&&
 &   f^{(c k \vec p)}(t,\vec x)=\f{1}{\sqrt{N_T V}}e^{i\omega^{(c k)} t
      + i\vec p \cdot\vec x}\,. 
  \end{aligned}
\end{equation}
Here and throughout the paper the eigenvalues are expressed in lattice
units. The spacetime dependence is fully contained in the plane
waves $f^{(c k \vec p)}(t,\vec x)$, while the spin and colour
dependence are encoded in the bispinors $\chi_\alpha^{(c k \vec p
  s\pm)}$, $\alpha=1,\ldots,4$, with $s=1,2$, and in the colour
vectors $\varphi_j^{(c)}$, $j=1,2,3$, with $c=1,2,3$, respectively.
The spatial momenta satisfy $\f{L}{2\pi}p_j=0,1,\ldots,L-1$, $L$ being
the spatial linear size of the lattice ($V=L^3$), to fulfill the
spatial periodic boundary conditions, while   
 \begin{equation}
   \label{eq:eff_Mats}
\omega^{(c k)} = \textstyle\f{1}{N_T}(\pi 
+  \phi_c + 2\pi k)=aT(\pi 
+  \phi_c + 2\pi k)  \,, \quad k=0,\ldots,N_T-1\,,
 \end{equation}
where $a$ is the lattice spacing and $N_T$ the temporal size of the
lattice, to fulfill the effective temporal boundary conditions. We
will refer to $\omega^{(c k)}$ as the effective Matsubara
frequencies. For a given value of the phase $\phi_c$, there are $N_T$
different branches for $\omega^{(c k)}$, corresponding, 
however, to only $N_T/2$ different eigenvalues, since the effective
Matsubara frequencies obey the relation
\begin{equation}
  \label{eq:mats_rel}
  \sin \omega^{(c k')} = -\sin \omega^{(ck)} ~~\text{if}~~
  k'=\textstyle\f{N_T}{2}+k ~{\rm mod}~ N_T\,. 
\end{equation}
Notice that the lowest positive branch of eigenvalues is described by
the function 
\begin{equation}
  \label{eq:lowst_br}
M(\phi_c)=\sin\f{\pi - |\phi_c|}{N_T}\,,  \quad \phi_c\in[-\pi,\pi]\,,
\qquad M(\phi_c+2\pi)=M(\phi_c)\,,
\end{equation}
which decreases as $\phi_c$ moves away from zero, where it is maximal,
and it is minimal when $e^{i\phi_c}=-1$. The bispinors
$\chi_\alpha^{(c k \vec p s\pm)}$, $s=1,2$, satisfy the equation 
\begin{equation}
  \label{eq:bisp}
  \begin{aligned}
    \f{\gamma_4 \sin\omega^{(c k)} + \sum_{j=1}^3 \gamma_j
      \sin p_j}{\sqrt{\sin^2\omega^{(c k)} + 
        \textstyle\sum_{j=1}^3\sin^2 p_j}}\, \chi^{(c k \vec p s\pm)} 
    &= \pm \chi^{(c k \vec p s\pm)} \,, \quad s=1,2\\
    \chi^{(c k \vec p s\pm)} {}^\dag\chi^{(c k \vec p s'\pm)}
    &=\delta_{ss'}\,,
  \end{aligned}
\end{equation}
where $\gamma_\mu$, $\mu=1,\ldots,4$ are Euclidean Dirac matrices. 
The spin index $s$ labels eigenmodes corresponding to the same
eigenvalue $\pm i\lambda^{(c k \vec p)}$. Taking into account the
degeneracy with respect to $k$ mentioned above, the eigenvalues are
therefore fourfold degenerate. Finally, the colour vectors are
trivially $\varphi_j^{(c)}=\delta_{cj}$, $c,j=1,2,3$, and satisfy 
$P\varphi^{(c)}=e^{i\phi_c}\varphi^{(c)}$.   

Let us now consider the qualitative features of the eigenmodes in the
general case. Above $T_c$, a typical gauge configuration consists of a
``sea'' where the Polyakov line $P({\vec x})$ gets ordered around
$\mathbf{1}={\rm diag}(1,1,1)$, which percolates through the lattice,
with ``islands'' of  ``wrong'' $P({\vec x})\neq \mathbf{1}$. If
$P({\vec x})$ were completely ordered, and spatial links were trivial,
there would be a sharp gap in the spectrum at $\lambda_g=M(0)=\sin(
\pi aT)$ (and a symmetric one at $-\lambda_g$), corresponding to the
lowest positive branch at $\phi_c=0$ in Eq.~\eqref{eq:eigen_simple}
(see also Eq.~\eqref{eq:lowst_br}). This is a ``zeroth-order''
picture, which is modified by the fluctuations of $P(x)$ and of the
spatial links. The ``first-order'' picture is obtained by allowing
$P(x)$ to fluctuate while still keeping the spatial links trivial. It
is clear from Eq.~\eqref{eq:lowst_br} that the quark eigenfunctions
can exploit the ``islands'' of ``wrong'' $P(x)$, which are
``energetically'' favourable, to lower their eigenvalues. In
particular, localising the wavefunction on the ``islands'' can achieve
a large eigenvalue reduction, if the momentum required to localise the
state is not too large, which is the case, for example, if the
``islands'' are sufficiently big.\footnote{  
  This can be seen by considering a configuration with 
  uniform ``islands'' of ``wrong'' Polyakov lines, and a test
  wavefunction of the form given in Eq.~\eqref{eq:eigen_simple} but
  localised and constant on one of such ``islands''. More precisely,
  take $\psi_{\rm test}(t,\vec x)=\psi^{(c k \vec 0 s+)}(t,\vec
  x)C_{\rm isl}(\vec x)$, with $k$ chosen so that $\lambda^{(c k \vec
    0)}$ belongs to the lowest positive branch. Here $C_{\rm isl}(\vec
  x)$ is 1 on the given ``island'' and 0 elsewhere. Computing the
  expectation value of $-\slaD^2$, one finds a value smaller than
  $\lambda_g^2$ if the ``island'' is big enough, so that surface
  effects at the interface with the ``sea'' do not overbalance the
  negative difference $\lambda^{(c k \vec 0 )}{}^2-\lambda_g^2$.} 
Therefore, in the ``first-order'' picture both localised and
delocalised eigenmodes appear between $-\lambda_g$ and $\lambda_g$,
with a few localised low modes well separated from the bulk of
delocalised modes. The full, ``second-order'' picture is finally
obtained by switching on the fluctuations of the spatial links. These
fluctuations have most likely a delocalising effect: for example, they
allow different colour components to mix (recall that we are working
in the diagonal temporal gauge). Nevertheless, the lowest modes can
still remain localised, due to the large energy difference with the
bulk of delocalised states. The sharp gap of the ``zeroth-order''
picture has turned into an ``effective gap'' $\lambda_c<\lambda_g$,
which we identify with the mobility edge separating localised and
delocalised modes. 
 
To better understand this picture, it is useful to set up an analogy 
between the Dirac operator and the Hamiltonian of an electron in a
crystal in the tight-binding approximation (TBA). If one discards the
spatial links, setting them to zero, the solutions of the Dirac
equation are localised at a given spatial site, and are given by
Eq.~\eqref{eq:eigen_simple} with $\vec p=0$, using the local Polyakov
line as the gauge background. These ``free'' solutions correspond to
the atomic orbitals in the TBA, which are localised on a crystal
site. In turn, the effect of the spatial links corresponds to that of
the hopping terms in the TBA, which allow the electron to hop between
crystal sites. In this analogy, the ``islands'' of ``wrong'' Polyakov
lines correspond to ``defects'' in the crystal (i.e., the diagonal
disorder in the Anderson model), and they act in the same way as
localising ``traps'' for the eigenmodes. Given the three-dimensional
nature of these ``islands'', the qualitative picture described above
allows one to understand why, in the high-temperature regime, the
critical properties of the Dirac spectrum at the
localisation/delocalisation transition correspond to those of the
three-dimensional unitary Anderson model. In order to check the
viability of our explanation, in the following Section we study an
effective model which should display localisation precisely through
the mechanism that we have proposed. In the remaining part of this
Section we add a few comments. 

The dependence of $\lambda_c$ on the temperature and on the lattice
spacing is expected to correspond qualitatively to that of
$\lambda_g$, which should somehow ``drag'' the mobility edge along. In
the ``zeroth-order'' picture, $\lambda_g=\sin\f{\pi}{N_T}=\sin (\pi
aT)$ increases when increasing the physical temperature at fixed $a$,
and it decreases as the lattice spacing is decreased at fixed physical 
temperature. This behaviour is qualitatively the same as that of the
Polyakov-loop expectation value, which is a measure of the ordering of
the Polyakov lines in the gauge configurations. Therefore, even though
$\lambda_g$ is determined assuming a perfectly ordered system, we can
loosely say that it responds in a direct way to the ordering of the
Polyakov lines. In the ``first-order'' picture, the separation point
$\lambda_g'$ between localised and delocalised modes lies below
$\lambda_g$. For a delocalised mode with low spatial momentum, which
sees all the fluctuations of the Polyakov lines equally, the gain in
``energy'' will average out, so bringing $\lambda_g'$ near the average
$\la M(\phi_c)\ra$ of the lowest positive branch. In the
``second-order'' picture, $\lambda_g'$ will be displaced further down 
to $\lambda_c$ by the mode mixing, but the size of this displacement
depends on the effectiveness of the mixing, which is difficult to
assess. It is likely, however, that an increase in the ordering of the
Polyakov lines will also make more correlated the spatial links at a
given spatial site but on different time slices. The fluctuations of
the spatial hopping should therefore become less pronounced, which in
turn should reduce the effectiveness of the mixing, and so reduce the
distance between $\lambda_c$ and $\lambda_g'$. In this case, both the
``first-order'' and the ``second-order'' effects tend to reduce the
distance between $\lambda_c$ and $\lambda_g$ as the system is made
more ordered, and to increase it when the ordering is reduced, thus
acting in the same direction as the ``zeroth-order'' effect. The
bottom line is that the response of $\lambda_c$ to a change in the
ordering of the system is expected to be qualitatively the same as
that of $\lambda_g$, i.e., it goes up with $T$ at fixed $a$, and it
goes down with $a$ at fixed $T$. This expectation is matched by the
results of numerical simulations~\cite{KP2}. 

The qualitative correctness of the ``zeroth-order'' picture 
allows also a qualitative understanding of the dependence of the
(pseudo)critical temperature $T_c$ on an imaginary chemical potential. 
As we have said above, in the high-temperature phase where the
Polyakov line gets ordered, the position of the mobility edge,
$\lambda_c$, is affected by the effective Matsubara frequency
corresponding to the trivial Polyakov line. At $\phi_c=0$, the
``twist'' of the quark wavefunction required by the effective boundary
conditions is maximal, and so the quantity $M(\phi_c)$, defined in
Eq.~\eqref{eq:lowst_br}, which ``drags'' $\lambda_c$, is maximal.   
Introducing an imaginary chemical potential $\mu$ (small enough in
order to avoid the Roberge-Weiss critical line~\cite{RW}) is
equivalent to adding an extra phase $e^{i\mu N_T}$ to the effective
boundary conditions, which reduces the ``twist''. As a consequence,
$M(\phi_c)\to M(\phi_c+\mu N_T)$, and in particular $M(0)\to M(\mu
N_T)$, i.e., it diminishes (independently of the sign of $\mu$), and
so one expects $\lambda_c$ to become smaller.\footnote{It is possible
  to argue that the expected dependence of $\lambda_c$ on $\mu$ is the
  same also in the ``first-order'' picture. The ``first-order''
  correction changes the sharp gap $\lambda_g$ into an effective gap
  $\lambda_g'<\lambda_g$, corresponding to the mobility edge in the
  ``first-order'' picture. As we discussed above, $\lambda_g'$ should
  be close to $\la M(\phi)\ra$, or, at nonzero $\mu$, to  $O(\mu)= \la
  M(\phi+\mu N_T)\ra$. The probability distribution function
  $f_{\mu}(\phi)$ of the Polyakov-line phases is a  periodic function, and 
  $f_{-\mu}(-\phi)=f_{\mu}(\phi)$ due to charge-conjugation
  invariance. This implies $O(\mu)=O(-\mu)$. In the ordered phase,
  even at nonzero $\mu$, $f_{\mu}(\phi)$ is peaked around zero,
  approximately symmetric, i.e., $f_{\mu}(-\phi)=f_{\mu}(\phi)$, and
  most likely monotonically decreasing in the interval
  $[0,\pi]$.  Exploiting these properties, one can show that $O(\mu)$ 
   is a decreasing function of $|\mu|$. 
}  If the relation between the chiral transition and the 
appearence of localised modes holds true, then one expects $T_c$ to
{\it increase} with $|\mu|$. This is actually what has been observed
in lattice simulations~\cite{immu,immu2}. 
The presence of the Roberge-Weiss critical lines can also
be understood in this framework, noticing that the system prefers to
have an effective gap as large as possible due to the fermionic
determinant. Comparing $M(\f{2\pi}{3}k+\mu N_T)$ for $k=-1,0,1$,
corresponding to the Polyakov line sectors $P=e^{i\f{2\pi}{3}k}$, one
finds that it becomes ``energetically'' favourable to switch 
to the sector $k=-1$ when $\mu N_T$ exceeds $\f{\pi}{3}$,
and similarly to the $k=1$ sector, and again to the trivial sector,
when $\mu N_T$ exceeds $\pi$ and $\f{5\pi}{3}$, respectively.  

As one would expect, our qualitative picture cannot explain all 
the features of the localised modes, especially when there are
competing effects at play. As it has been shown in
Ref.~\cite{KP2}, the lowest-lying Dirac eigenmodes should remain
localised also in the continuum limit. While $\lambda_c$ goes to 
zero as $a\to 0$, in agreement with our qualitative explanation, the
``renormalised mobility edge'', $\lambda_c/m_{ud}$, with $m_{ud}$ the
bare light-quark mass, has a finite continuum limit, and the number of
localised modes per unit volume, $n_{\rm loc}$, 
\begin{equation}
  \label{eq:nloc}
n_{\rm loc}=\f{1}{Va^3}\int_0^{\lambda_c}d\lambda\,\rho(\lambda)\,,
\end{equation}
also remains finite in the continuum. Here $\rho(\lambda)$ is the
spectral density in lattice units. Within our picture, the fact that
localisation survives the continuum limit is the nontrivial outcome of
two competing effects. In fact, while the ordering of the Polyakov
lines tends to disappear as $a$ is reduced, thus lowering $\lambda_c$, 
at the same time $\rho(\lambda)$ at the low end of the spectrum is
increased, due to the larger amount of fluctuations, and so of
``islands'' that can support localised modes. In Ref.~\cite{KP2} it
has also been observed that $n_{\rm loc}$ increases rapidly with
$T$. At fixed lattice spacing, increasing the temperature brings
$\lambda_c$ up, while at the same time decreasing the spectral
density. Both phenomena are in agreement with our explanation, since
increasing the ordering of the Polyakov lines, besides pushing up the
mobility edge, also reduces the density of fluctuations and thus
the spectral density of localised modes.\footnote{
  Notice that since the Polyakov-line phases are continuous variables,
  it is possible to have a finite amount of ``islands'' also at very
  high temperature, as their ``energy'' cost can be contained by making
  their deviation from the ordered value arbitrarily small.} 
The value of $n_{\rm loc}$ at fixed $a$ results from the balance of
these two effects. On top of this, to determine the physical value of
$n_{\rm loc}$ one has to take the continuum limit, which has an effect
on $\lambda_c$ and $\rho(\lambda)$ opposite to that of increasing the
temperature, as we have said above. The actual behaviour of $n_{\rm
  loc}$ as a function of the temperature ultimately depends on the
relative magnitude of all these effects, which cannot be determined in
our simple picture.

\section{Ising-Anderson model}
\label{sec:IA}

The considerations of the previous Section suggest that it should be
possible to understand the qualitative features of the Dirac spectrum
and eigenfunctions in QCD, for what concerns the localisation
properties, by using a genuinely three-dimensional model. To construct
such a model one has to strip off all the features that are irrelevant
to localisation.  

The first step is to get rid of the time direction, reducing the
lattice to three dimensions, and replacing the time covariant
derivative in the Dirac operator with a  diagonal noise term, intended
to mimic the effective boundary conditions. This is motivated by the
strong correlation among time-slices, as already mentioned in the
previous Section. 

Furthermore, it is known that, in general, off-diagonal disorder is
less effective than diagonal disorder in producing
localisation~\cite{offdiag,offdiag2}, so it should be safe to replace
spatial covariant derivatives with ordinary derivatives.  
More precisely, in our case the structure of the hopping terms is of
the form considered in Ref.~\cite{GGC}. It is shown there that the
width of the disorder distribution has to be rather large to produce
localised modes near $\lambda=0$ through off-diagonal disorder
only. This is definitely not the case in QCD, where the disorder in
the hopping terms involves unitary matrices. As we have already
remarked, fluctuations of the spatial links have most likely a
delocalising effect on the low modes. While this is certainly
important for the detailed features of the spectrum and of the
eigenmodes observed in QCD, in our picture it can nevertheless be
regarded as a ``second-order'' effect, acting on the localised modes
which should be produced by the ``first-order'' effect, i.e., the
fluctuations of the Polyakov lines. To be precise, it is thus the
``first-order'' picture, discussed in the previous Section, that we
are going to test here. Notice that replacing the spatial covariant
derivatives with ordinary derivatives decouples the colour components
of the quark wavefunction: this changes the symmetry class (in the
sense of random matrix models), as we discuss in detail below. 

If our ``sea/islands'' mechanism is viable, then the main features of
localisation should still be captured by a genuine three-dimensional
model of the following general form, 
\begin{equation}
  \label{eq:Heff}
  H^{\rm eff}_{\vec x \vec y} = \gamma_4 {\cal N}_{\vec x}\delta_{\vec x
    \vec y} + i\vec\gamma \cdot \vec\de_{\vec x    \vec y}\,,
  \qquad (\de_j)_{\vec x \vec y} =
  {\textstyle\f{1}{2}}(\delta_{\vec x+\hat 
    \jmath,\vec y}-\delta_{\vec x-\hat \jmath,\vec y})\,,
\end{equation}
if the diagonal noise term ${\cal N}_{\vec x}$ has the same features
as the ``diagonal noise'', i.e., the effective boundary conditions,
appearing in the true ``Hamiltonian'' $-i\slaD$. The relevant features
are the following: 
\begin{itemize}
\item the diagonal noise involves continuous variables, namely the
  phases of the local Polyakov lines;
\item the diagonal noise terms are correlated, and are governed by the
  dynamics of the local Polyakov lines;
\item as the system is made more ordered, the diagonal noise tends to
  introduce a gap in the spectrum.
\end{itemize}
The simplest way to incorporate these features in ${\cal N}_{\vec x}$  
is to base the diagonal noise on a spin model with continuous spins:
this obviously satisfies the first requirement, and also the second
one as Polyakov lines in the high-temperature phase display indeed a
spin-model type dynamics~\cite{Yaffe:1982qf,DeGrand:1983fk}. Finally,
also the third requirement is easily implemented if we take
\begin{equation}
  \label{eq:noise}
  {\cal N}_{\vec x} =\Lambda\f{1+s_{\vec x}}{2}\,, \qquad s_{\vec
    x}\in[-1,1]\,, 
\end{equation}
where $s_{\vec x}$ is the spin variable at point $\vec x$, and
$\Lambda$ is a constant determining the strength of the coupling of
the fermions to the spins. The distribution of the spins $s_{\vec x}$
is determined by the dynamics of an Ising-like model with continuous
spins, which we take of the simplest possible form with
nearest-neighbour interactions only, 
\begin{equation}
  \label{eq:ising}
  \f{H_{\rm Ising}}{kT} = -\beta_{\rm Ising} \sum_{<\vec x\,\vec y>}
  s_{\vec x} s_{\vec y}\,, \qquad s_{\vec x}\in[-1,1]\,.
\end{equation}
Since $\frac{1+s_{\vec x}}{2}$ is 1 for ``aligned'' spins (i.e., for
$s_{\vec x}=1$), this choice provides indeed an effective spectral
gap when the spins are ordered.\footnote{It is understood that we work
  with magnetic field $h=0^+$.} 
In the ordered phase of the Ising model there is a ``sea'' of $s_{\vec
  x}\simeq 1$ spins with ``islands'' of $s_{\vec  x}\neq 1$ spins, so the
underlying configurations have the same features as the Polyakov line 
configurations in QCD. In QCD we have a single parameter
governing the ordering of the configuration and the size of the
effective gap. In the effective model the ordering of the spin
configuration is governed by $\beta_{\rm Ising}$, while the size of
the gap is mainly determined by the spin-fermion coupling $\Lambda$,
although one expects it to be affected also by the magnetisation of
the system, and thus by $\beta_{\rm Ising}$.

We notice in passing a curious feature of this model: it
belongs to different symmetry classes for lattices of even or odd size
$L$, namely to the orthogonal class for $L$ even, and to the
symplectic class for $L$ odd (see Appendix \ref{app:B}). For our
purposes it is convenient to work with even-sized lattices, which
allow to get rid of the Dirac degree of freedom through a spin
diagonalisation.  

Let us now describe the spectrum of our effective model in more
detail. The spectrum is symmetric with respect to the origin on a
configuration-by-configuration basis, due to the fact that
$\{\gamma_5,H^{\rm eff}\}=0$. In order to find the eigenvalues, it is
convenient to first perform a spin diagonalisation through the
following unitary transformation, 
\begin{equation}
  \label{eq:spin_diag}
  U_{\vec x \vec y}=\delta_{\vec x \vec
    y}\,\gamma_1^{x_1}\gamma_2^{x_2}\gamma_3^{x_3} 
\,,
\end{equation}
which yields
\begin{equation}
  \label{eq:Heff_sd}
  {\cal H}_{\vec x \vec y}  \equiv (U^\dag H^{\rm
    eff} U)_{\vec x \vec y}= \eta_4(\vec x) {\cal N}_{\vec
    x}\delta_{\vec x \vec y}\,\gamma_4 +  i\vec\eta(\vec x) \cdot \vec\de_{\vec x
    \vec y}\,\mathbf{1}\,, 
\end{equation}
where $\mathbf{1}$ is here the identity in Dirac space, and
\begin{equation}
  \label{eq:etas}
\eta_\mu(\vec x)=(-1)^{ \left(\sum_{\nu=1}^\mu x_\nu\right) -x_\mu
}\,. 
\end{equation}
The Hamiltonian ${\cal H}$  obviously commutes with $\gamma_4$, so one
can diagonalise them simultaneously. Denoting with $\psi_{\pm n}$ the
common eigenvectors, ${\cal H} \psi_{\pm n} = \lambda_{\pm n}\psi_{\pm
  n}$, $\gamma_4 \psi_{\pm n} = \pm \psi_{\pm n}$, one has for the
bispinors $\psi_{\pm n}$ 
\begin{equation}
  \label{eq:g4_ev}
  \psi_{\pm n}(\vec x) = \left(
    \begin{array}{c}
      \xi_{\pm n}(\vec x) \\
      \pm \xi_{\pm n}(\vec x)
    \end{array}
\right)\,, \qquad
  \xi_{\pm n}(\vec x) = \left(
    \begin{array}{c}
      \xi^{(1)}_{\pm n}(\vec x) \\
      \xi^{(2)}_{\pm n}(\vec x)
    \end{array}
\right)\,,
\end{equation}
and the identical single-component eigenvalue equations 
\begin{equation}
  \label{eq:g4_ev_H}
    {\cal H}^{(\pm)} \xi^{(s)}_{\pm n} 
    = \lambda_{\pm n}\, \xi^{(s)}_{\pm n}\,,\quad s=1,2\,,
\end{equation}
where
\begin{equation}
  \label{eq:g4_ev_Hbis}
    {\cal H}^{(\pm)}_{\vec x \vec y} = \pm\eta_4(\vec
    x) {\cal N}_{\vec x}\delta_{\vec x 
        \vec y} + i\vec\eta(\vec x) 
      \cdot \vec\de_{\vec x    \vec y}\,.
\end{equation}
It is immediate to notice that there is an exact twofold degeneracy of
the eigenvalues. It is also straightforward to prove that  ${\cal
  H}^{(-)} \eta_4=- \eta_4{\cal H}^{(+)}$, so setting $\xi^{(s)}_{-
  n}(\vec x) = \eta_4(\vec x)\xi^{(s)}_{+ n}(\vec x)$ one finds 
\begin{equation}
  \label{eq:g4_ev_H_2}
{\cal H}^{(-)}\xi^{(s)}_{- n} = -\lambda_{+ n} \xi^{(s)}_{- n}=
\lambda_{- n} \xi^{(s)}_{- n}\,, 
\end{equation}
i.e., $\xi^{(s)}_{- n}$ is an eigenvector of ${\cal H}^{(-)}$ with
eigenvalue $\lambda_{- n} = -\lambda_{+ n}$. This shows again that the
spectrum is symmetric with respect to the origin. 
 
It is less straightforward to show that for a given noise
configuration ${\cal N}_{\vec x}$ and a given eigenvector 
$\xi_{+}$ with eigenvalue $\lambda_+$, there is another 
configuration with noise $\bar{\cal N}_{\vec x}$ and with the same
Boltzmann weight, and an eigenvector $\bar\xi_{+}$ corresponding to
the eigenvalue $-\lambda_+$. Indeed, taking $\bar{\cal N}_{\vec x}$ and
$\bar\xi_{+}$ as follows, 
\begin{equation}
  \label{eq:appro_sym}
  \bar{\cal N}_{\vec x} = {\cal N}_{\vec x+ \hat 3}\,,\qquad
  \bar\xi_{+}(\vec x) = \eta_4(\vec x)\xi_{+}(\vec x+ \hat 3)\,,
\end{equation}
one can explicitly verify that this is the case. This implies that the
average spectrum of ${\cal H}^{(+)}$ (and so that of ${\cal
  H}^{(-)}$), obtained by integrating over the disorder (i.e., over
the spin configurations), has an exact symmetry with respect to the
origin.  

\section{Numerical results}
\label{sec:NR}

\begin{figure}[t]
  \centering
  \includegraphics[width=0.48\textwidth]{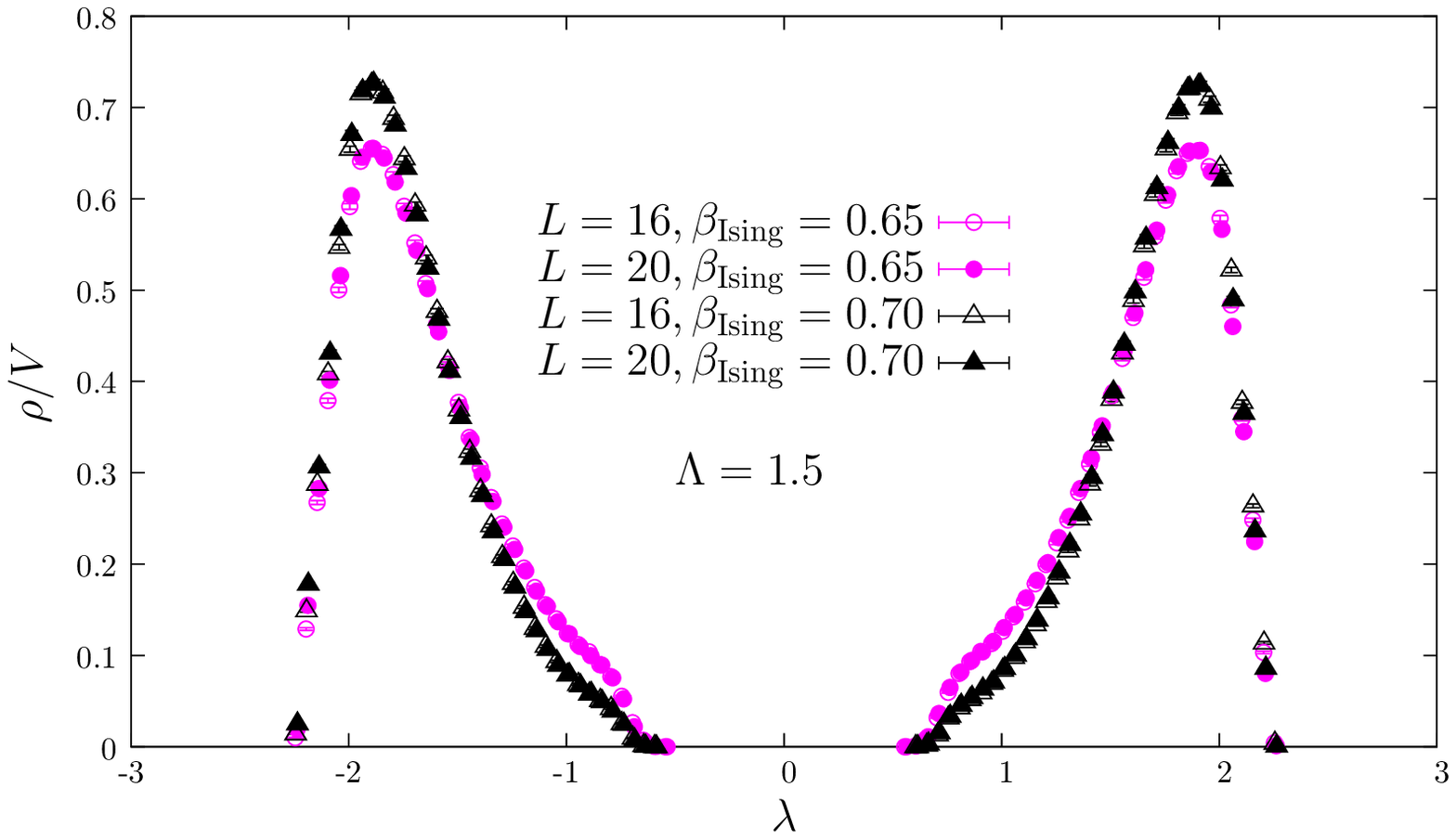}\hfil
  \includegraphics[width=0.48\textwidth]{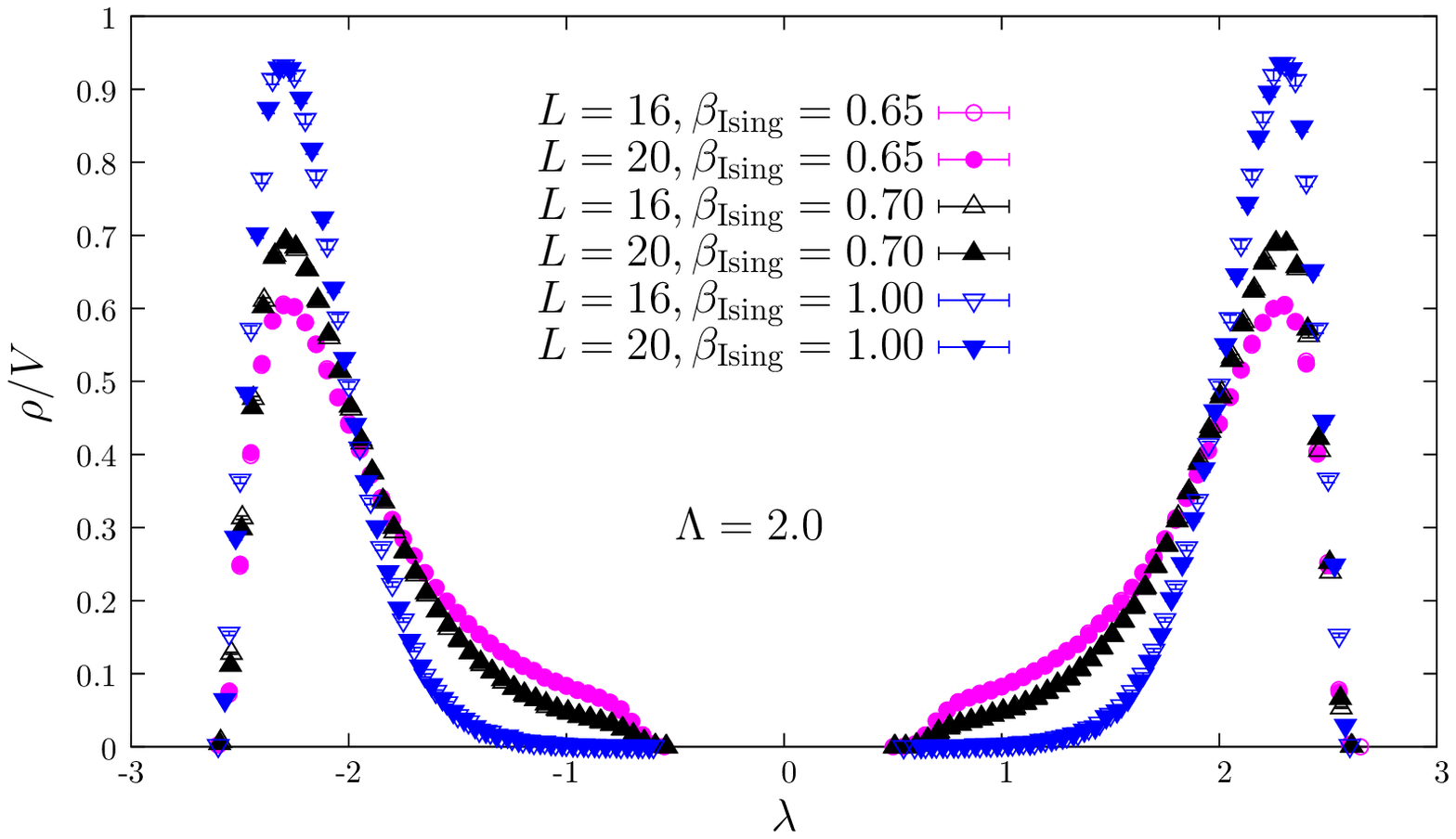}
  \caption{Spectral density of the Ising-Anderson model (reduced) 
    Hamiltonian ${\cal H}^{(+)}$, Eq.~\eqref{eq:g4_ev_Hbis}, for
    several choices of the parameters 
    $\beta_{\rm Ising}$ and $\Lambda$.
  }
  \label{fig:1}
\end{figure}

We have performed numerical simulations of the effective 
Ising-Anderson model discussed in the previous Section,
Eqs.~\eqref{eq:Heff}, \eqref{eq:noise} and \eqref{eq:ising}.  
For our purposes it was sufficient to focus on ${\cal H}^{(+)}$,
Eq.~\eqref{eq:g4_ev_Hbis}, which contains all the relevant information
on the localisation properties of the eigenmodes. To form an overall
picture of the properties of the effective model, we have performed
full diagonalisation of the Hamiltonian on medium-size lattices
($L=16$ and $L=20$), for inverse temperatures $\beta_{\rm
  Ising}=0.65,\,0.70$, in the ordered phase of the continuous-spin
Ising system, and spin-fermion coupling $\Lambda=1.5,\,2.0$, on 300
spin configurations. We have furthermore fully diagonalised ${\cal
  H}^{(+)}$ on lattices of the same size for $\beta_{\rm Ising}=1.0$,
well above the critical inverse temperature, setting $\Lambda=2.0$.  

In Fig.~\ref{fig:1} we show the spectral density per unit volume of
the system. Low modes have small spectral density, which rapidly
increases as one goes up in the spectrum, as expected. Decreasing the
temperature, thus making the system more ordered, decreases the
spectral density of low modes, as one expects if these modes are
localised on fluctuations of the spin variables. Increasing the
spin-fermion coupling enlargens the region where the spectral density
is small, again as expected, since it should push the effective gap up
in the spectrum. The same effect is obtained by increasing $\beta_{\rm
  Ising}$, i.e., by making the system more ordered. Notice the symmetry
under $\lambda \to -\lambda$ of the average spectrum, which has been
discussed in the previous Section. There is apparently also a sharp
gap in the spectrum, which is however not very sensitive to $\Lambda$,
$\beta_{\rm Ising}$, or to the size of the system. This gap is
probably a feature of the effective model, due to the drastic
simplifications of the spatial hopping terms; in any case, it is
irrelevant for our purposes.  

\begin{figure}[t]
  \centering
  \includegraphics[width=0.48\textwidth]{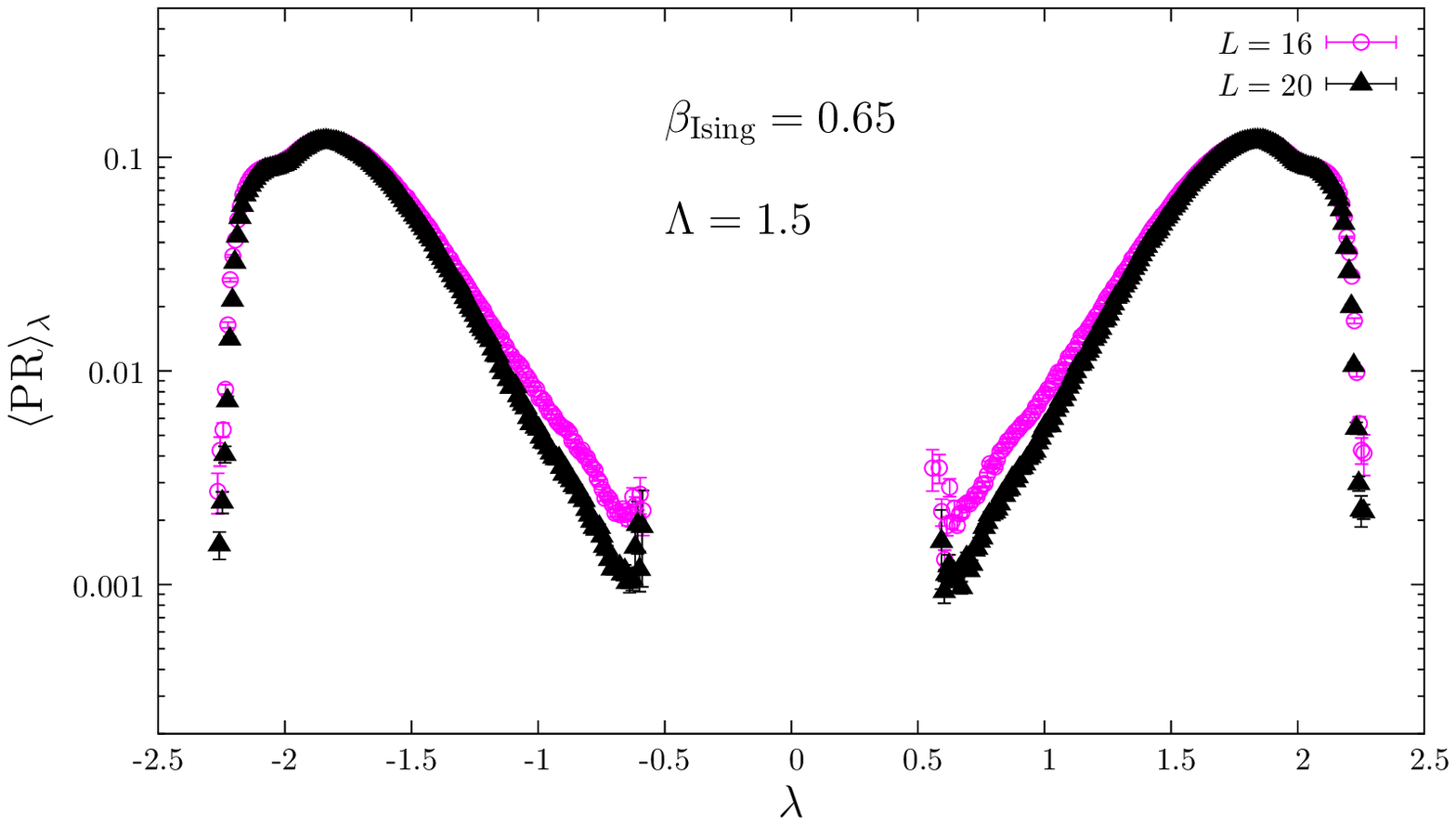}
  \includegraphics[width=0.48\textwidth]{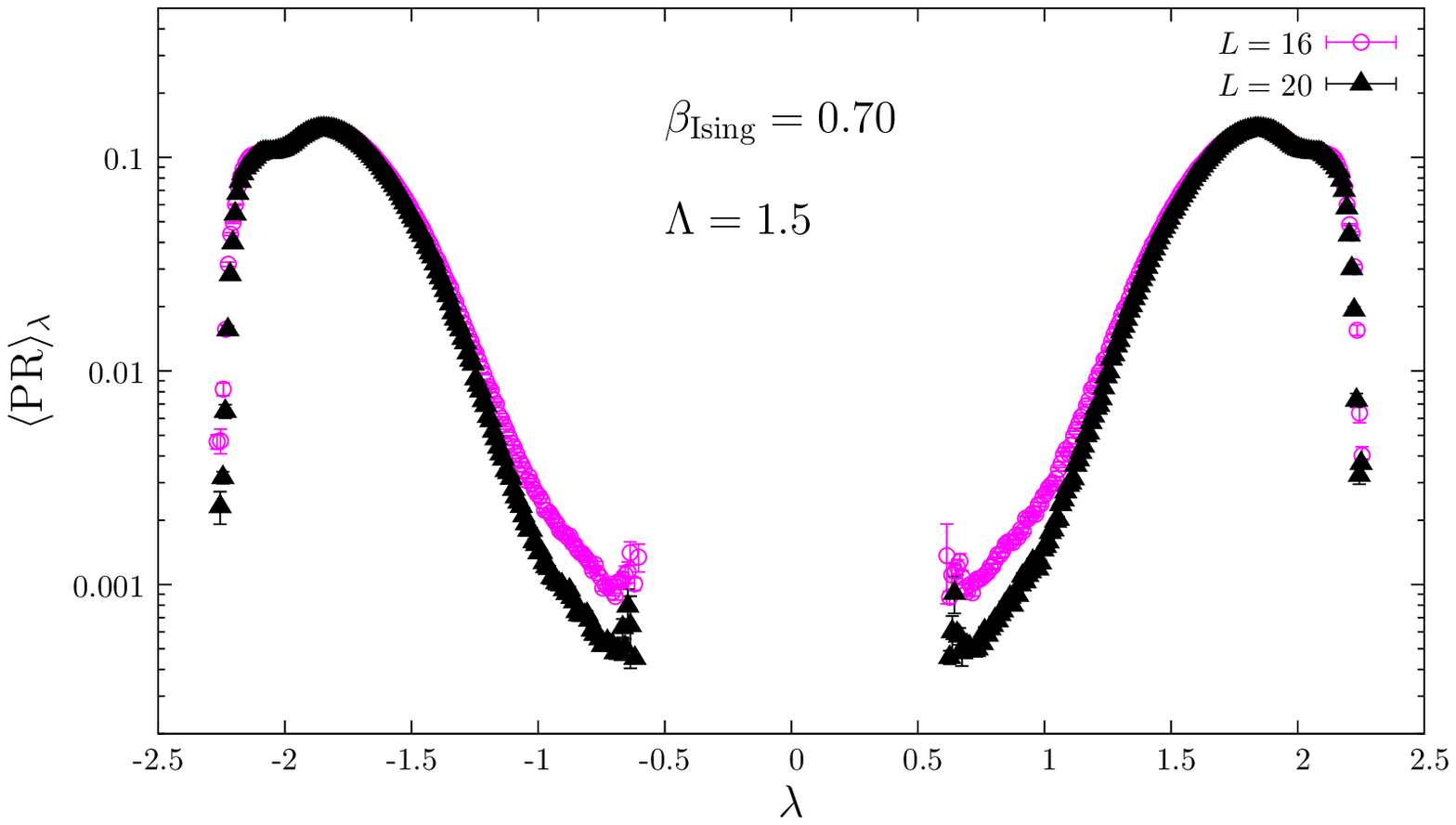}

  \includegraphics[width=0.48\textwidth]{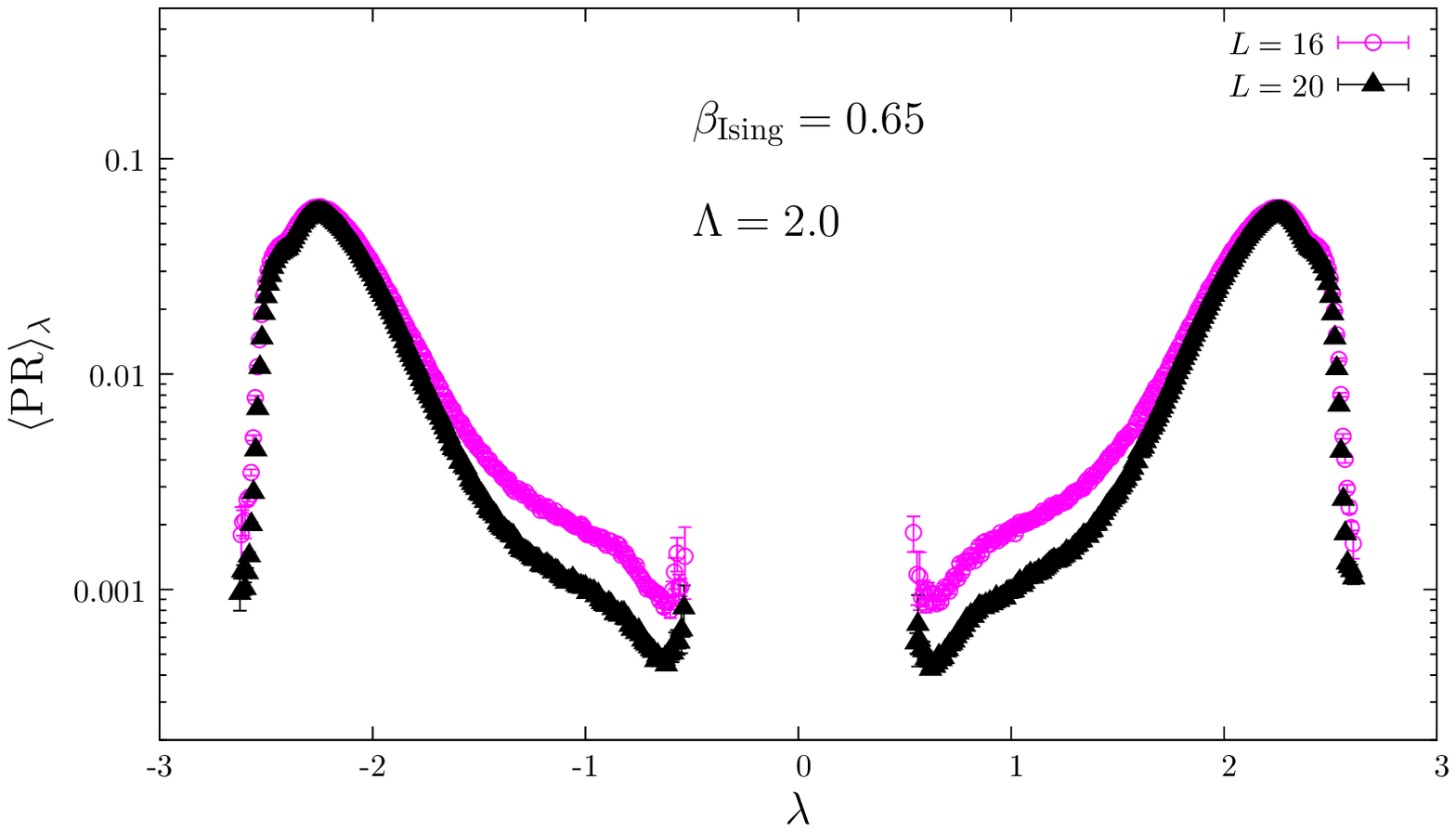}  
  \includegraphics[width=0.48\textwidth]{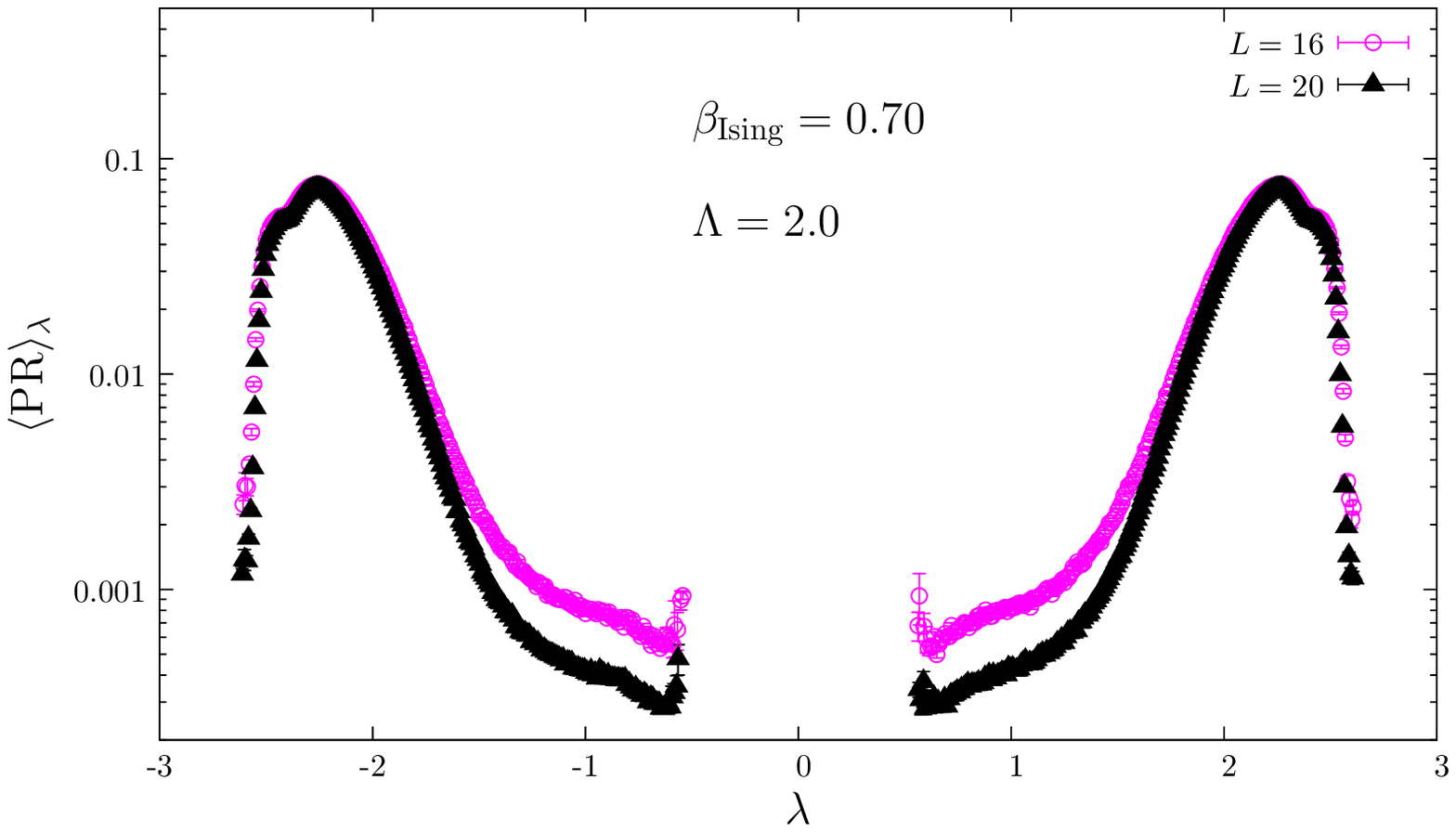}
  \caption{Participation ratio of eigenmodes of 
    the Ising-Anderson model (reduced) 
    Hamiltonian ${\cal H}^{(+)}$, Eq.~\eqref{eq:g4_ev_Hbis},
    for $\beta_{\rm Ising}=0.65,\,0.70$ and $\Lambda=1.5,\,2.0$.}
  \label{fig:2bis}
\end{figure}

\begin{figure}[t]
  \centering
  \includegraphics[width=0.48\textwidth]{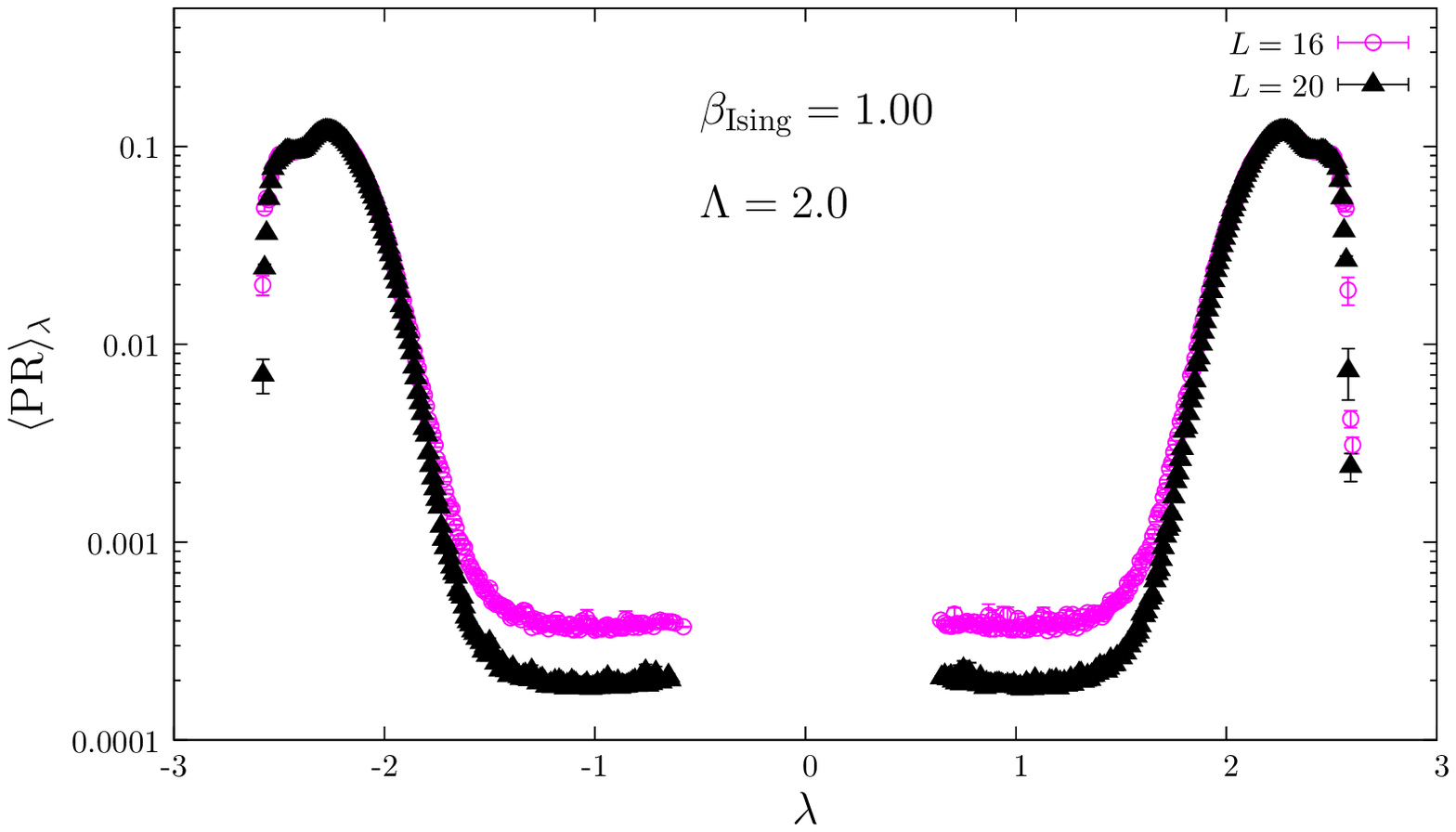}  
  \includegraphics[width=0.48\textwidth]{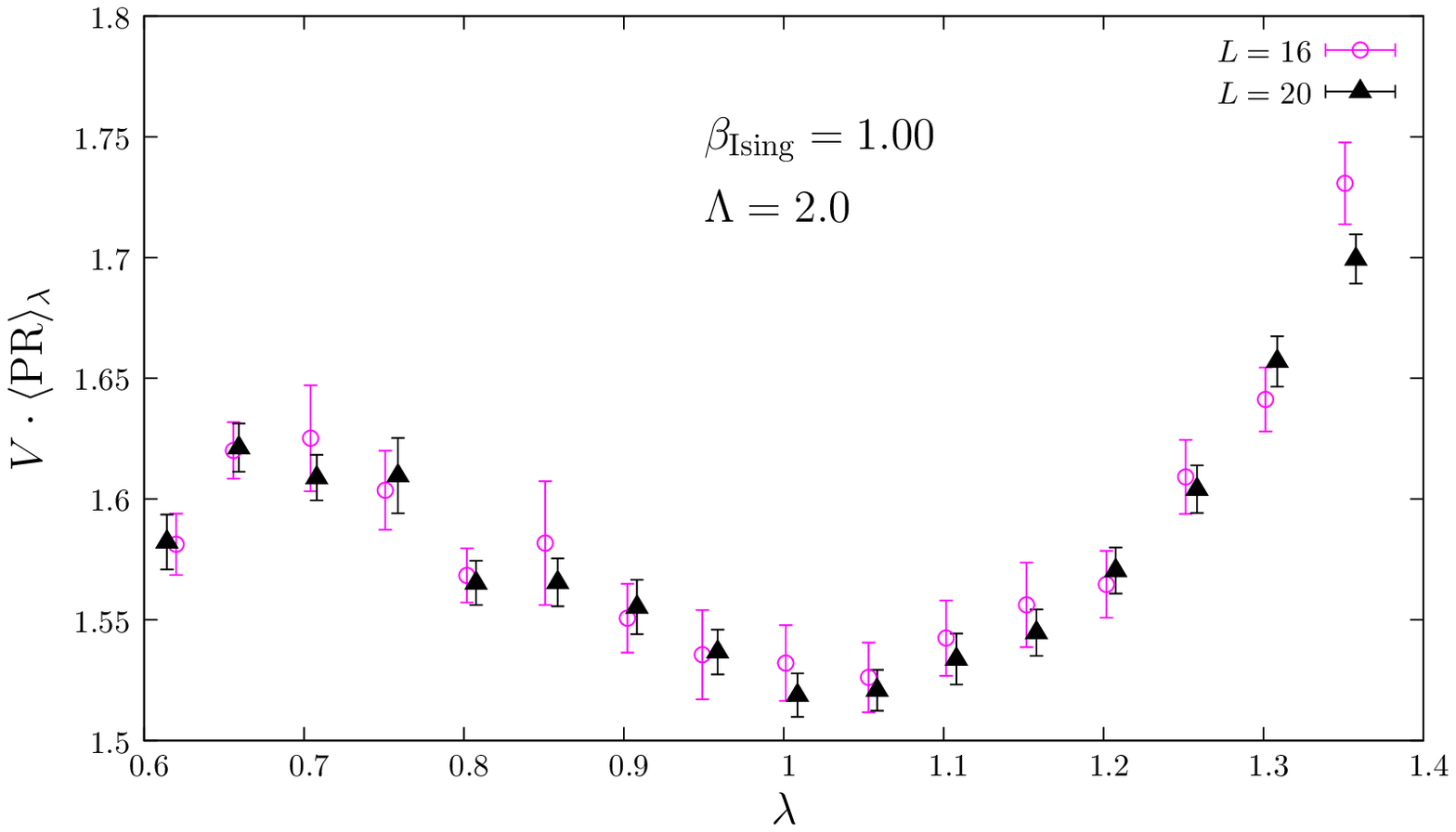}
  \caption{Participation ratio of eigenmodes of 
    the Ising-Anderson model (reduced) Hamiltonian ${\cal H}^{(+)}$,
    Eq.~\eqref{eq:g4_ev_Hbis}, for $\beta_{\rm Ising}=1.0$ and
    $\Lambda=2.0$. The right panel is a zoom on the low-lying modes.}  
  \label{fig:2ter}
\end{figure}

In Figs.~\ref{fig:2bis} and \ref{fig:2ter} we show the average
participation ratio ${\rm PR}=\la (\sum_x |\psi(x)|^4)^{-1}/
V\ra$, which gives a measure of the fraction of space occupied by a
given mode, as a function of its location in the
spectrum.\footnote{Normalisation to 1 of the eigenfunction
  $\psi(x)$ is understood.} The ${\rm PR}$ changes by two 
orders of magnitude as one moves up in the spectrum starting from the
low-density region; more importantly, when increasing the size of the
system it remains almost constant in the bulk of the spectrum, while
it visibly decreases near the origin. This signals that modes in the
bulk are delocalised, while modes near the origin are
localised. Notice in Fig.~\ref{fig:2ter} the upturn of the ${\rm PR}$ at the
low end of the spectrum: a similar phenomenon has been observed in QCD
with staggered fermions~\cite{KP2}.  

\begin{figure}[t]
  \centering
  \includegraphics[width=0.48\textwidth]{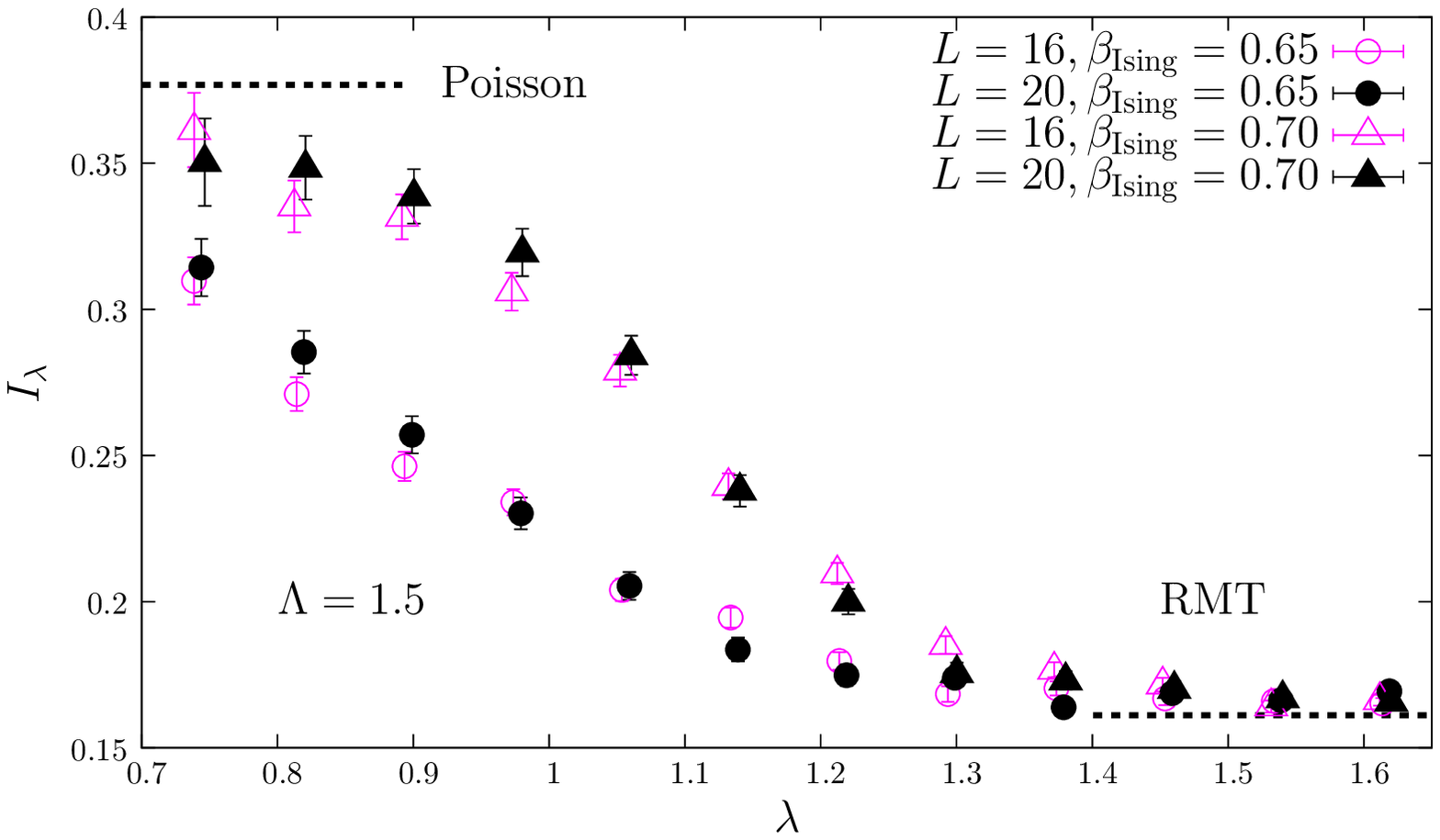}\hfil
  \includegraphics[width=0.48\textwidth]{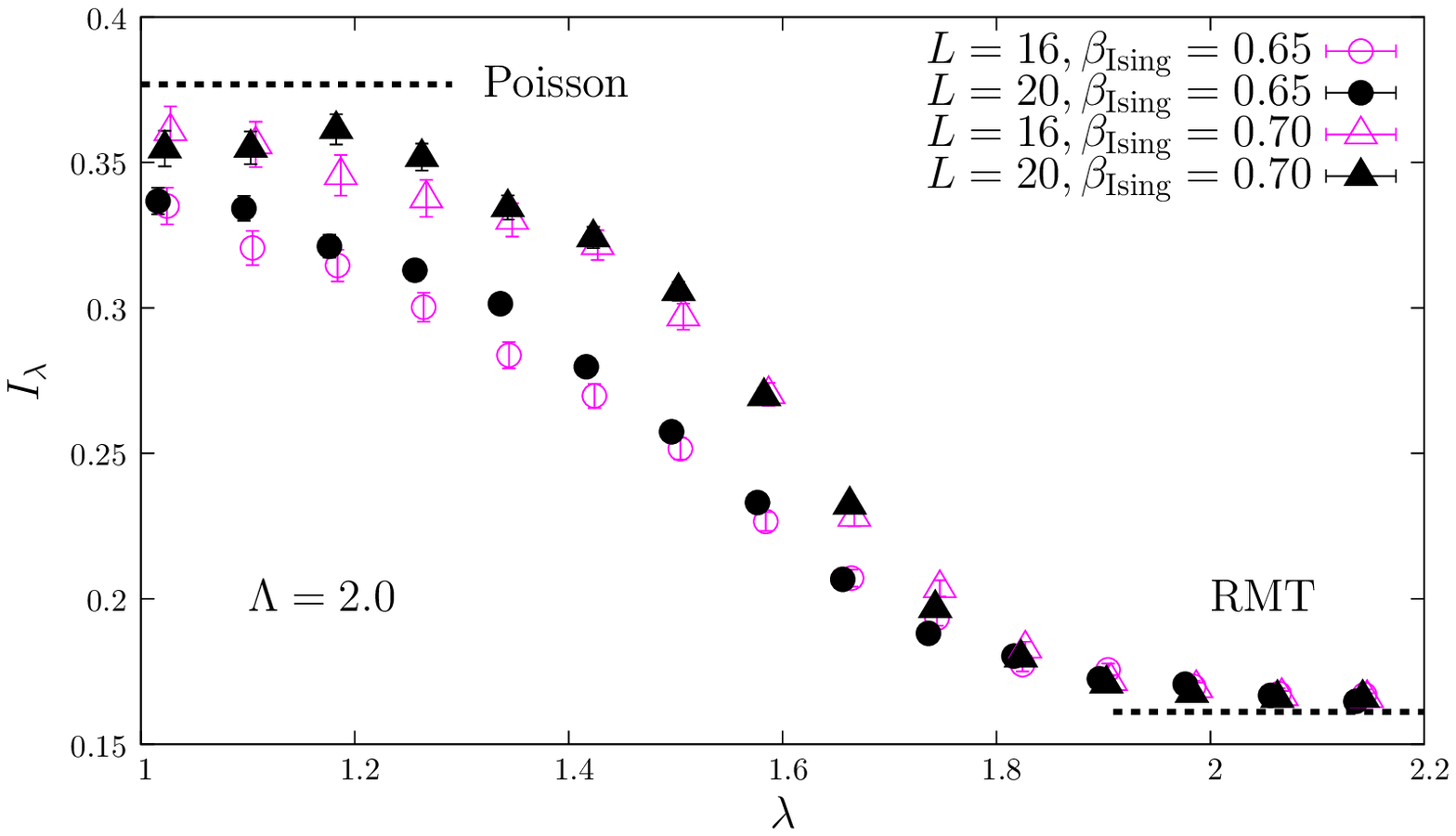}  
  \includegraphics[width=0.48\textwidth]{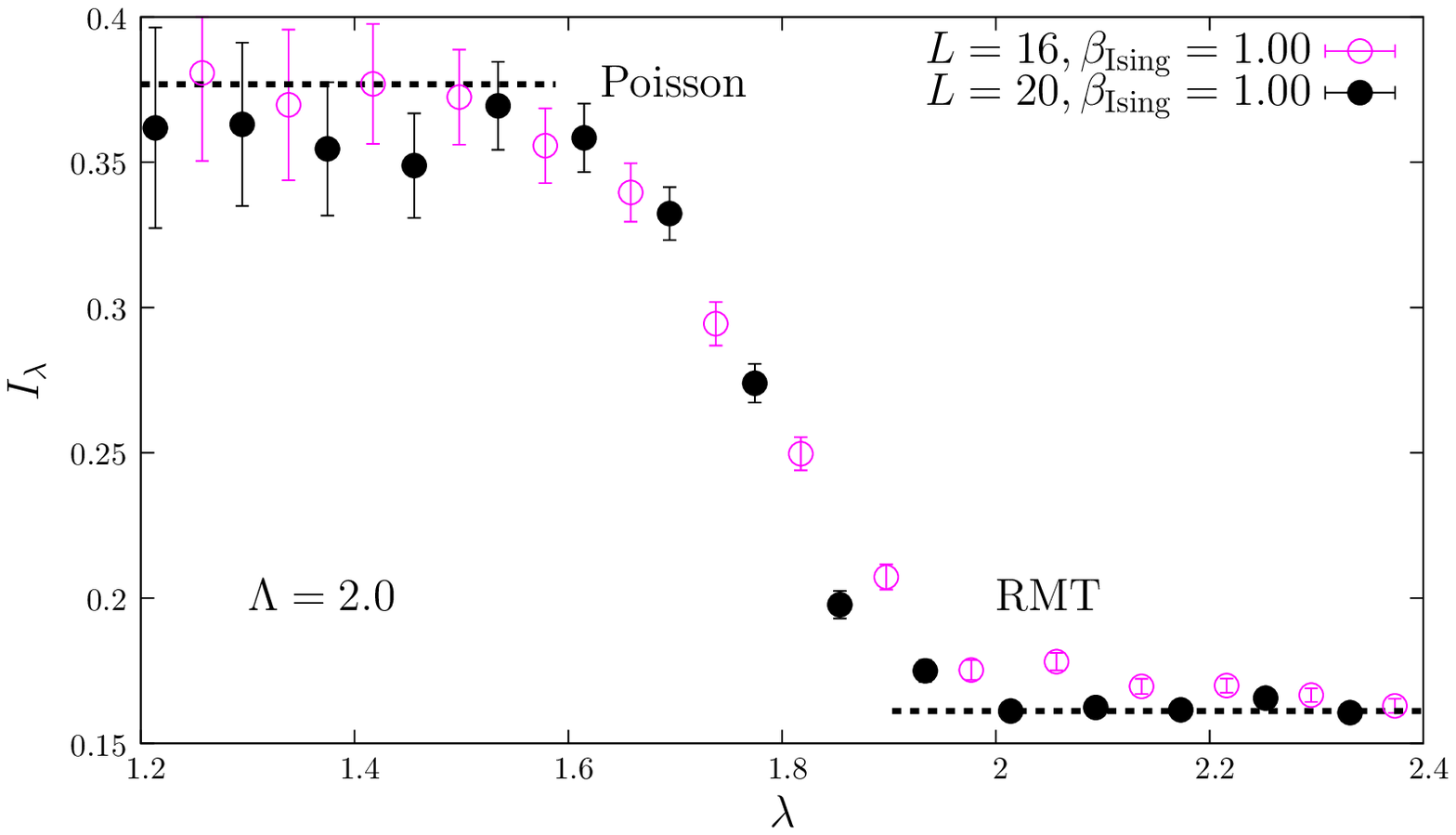}  
  \caption{The spectral statistics $I_\lambda$,
    Eq.~\protect\eqref{eq:ilam}, for the eigenmodes of 
    the Ising-Anderson model (reduced) 
    Hamiltonian ${\cal H}^{(+)}$, Eq.~\eqref{eq:g4_ev_Hbis},
    for several choices of $\beta_{\rm Ising}$ and $\Lambda$.
  } 
  \label{fig:3}
\end{figure}

Finally, in Fig.~\ref{fig:3} we show a suitably defined local spectral
statistics $I_\lambda$ across the spectrum. Spectral statistics can be
used to detect a localisation/delocalisation transition, since the
eigenvalues corresponding to localised or delocalised eigenmodes are
expected to obey different statistics, namely Poisson or Wigner-Dyson
statistics, respectively. More precisely, $I_\lambda$ is defined as  
\begin{equation}
  \label{eq:ilam}
I_\lambda = \int_0^{\bar s} ds\, p_\lambda(s)\,, \qquad
s_i=\f{\lambda_{i+1}-\lambda_i}{\la\lambda_{i+1}-\lambda_i\ra_\lambda}\,,\qquad 
\bar s\simeq 0.473\,,
\end{equation}
where the {\it unfolded level spacing distribution} (ULSD)
$p_\lambda(s)$ is the probability distribution, computed locally
in the spectrum, of the so-called {\it unfolded level spacing} $s_i$,
i.e., the level spacing divided by the local average level spacing
$\la\lambda_{i+1}-\lambda_i\ra_\lambda$. The quantity $I_\lambda$ is
simply the integrated ULSD up to the crossing point $\bar s$ of the
exponential distribution, corresponding to Poisson statistics,  and
the orthogonal Wigner surmise, which accurately describes the ULSD for
(orthogonal) Wigner-Dyson statistics. The results confirm that the
eigenmodes change from localised to delocalised when one moves up in
the spectrum. Moreover, the mobility edge separating localised and
delocalised modes goes up in the spectrum when the Ising system is
made more ordered by decreasing the temperature, or when the
spin-fermion coupling is increased. Our results give also a
first indication that the slope of the curve increases as the volume
is increased, thus hinting at the existence of a true phase transition
in the spectrum. Notice that the transition takes place near the point
$\Lambda\f{1+\la s_x\ra}{2}$ in the spectrum, thus indicating that the
position of the mobility edge is connected to the magnetisation of the
system, as expected (see Table \ref{tab:1}). This is also in agreement
with our expectation for the position of the mobility edge in the
``first-order'' picture, discussed previously in Section
\ref{sec:PL}. 

\begin{figure}[t]
  \centering
  \includegraphics[width=0.48\textwidth]{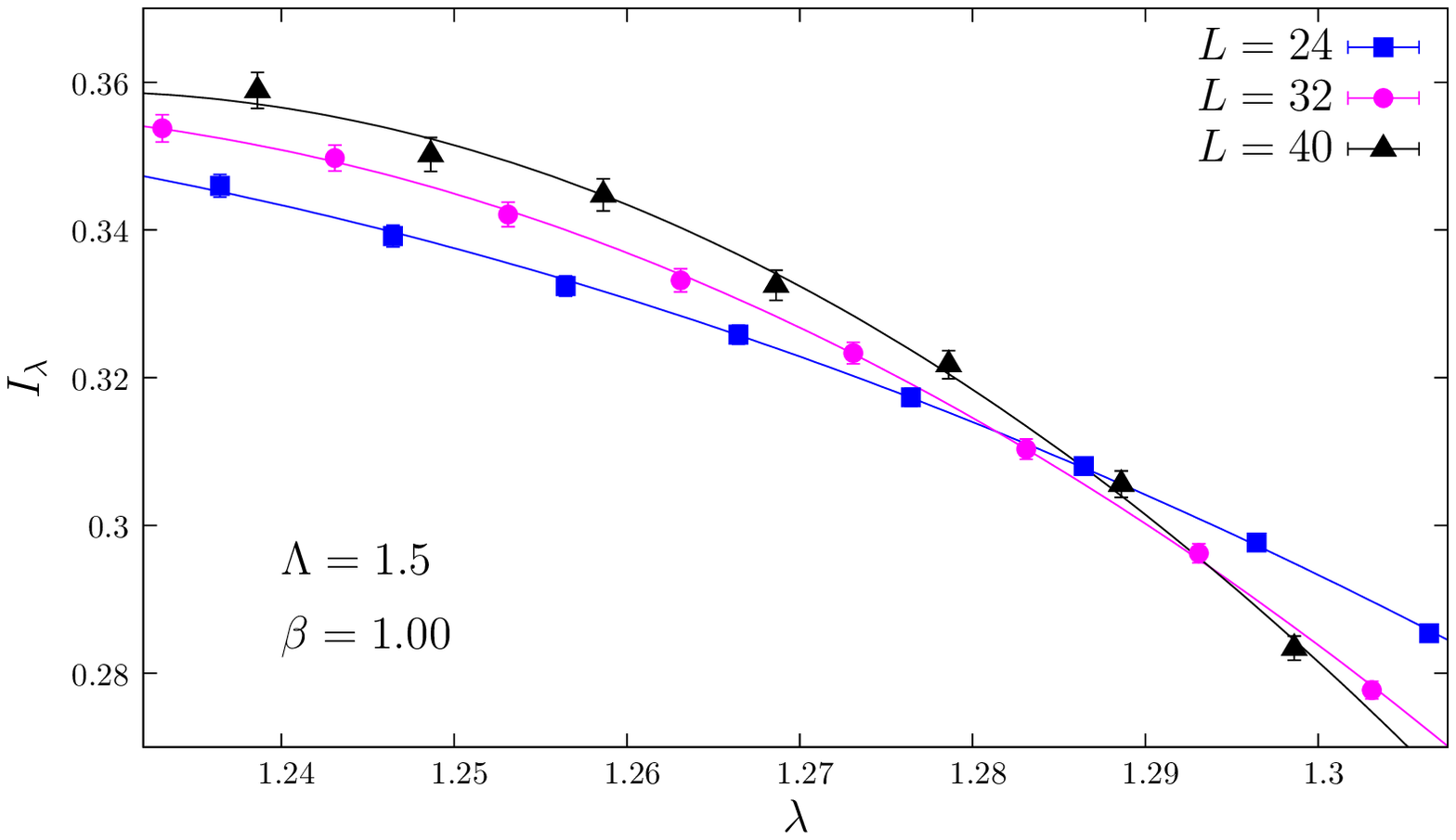}
  \includegraphics[width=0.48\textwidth]{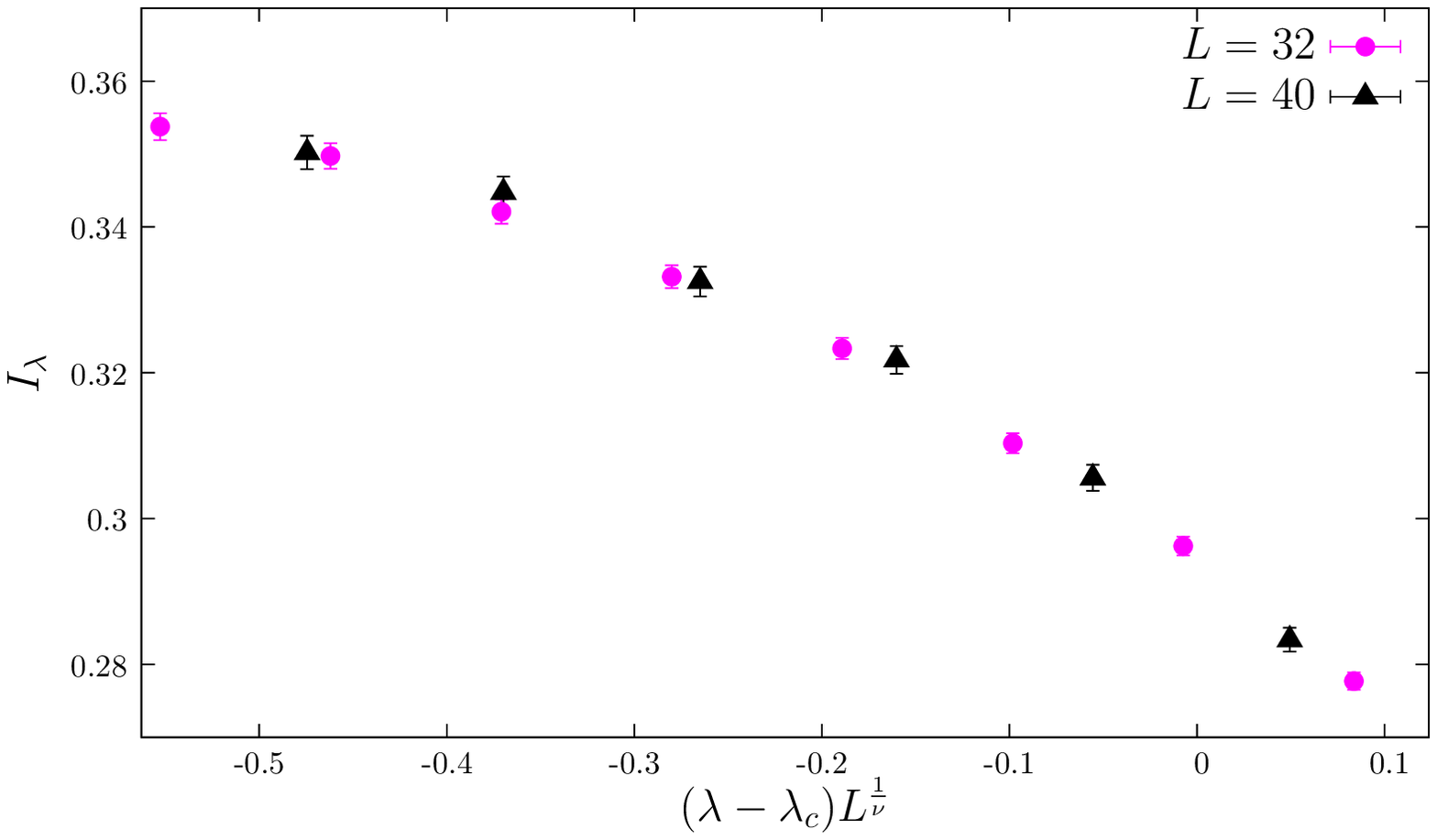}
  \caption{(Left panel) The spectral statistics $I_\lambda$,
    Eq.~\protect\eqref{eq:ilam}, for the eigenmodes of 
    the Ising-Anderson model (reduced) Hamiltonian ${\cal H}^{(+)}$,
    Eq.~\eqref{eq:g4_ev_Hbis}, near the critical point in the
    spectrum. Here  $\beta_{\rm Ising}=1.0$ and $\Lambda=1.5$. The
    solid line is a quadratic fit to the data, and is shown to guide
    the eye. (Right panel) The same spectral statistics for
    $L=32,\,40$ plotted against the scaling variable
    $(\lambda-\lambda_c)L^{\f{1}{\nu}}$.  
  } 
  \label{fig:3bis}
\end{figure}

To further investigate the onset of critical behaviour, we have
determined eigenvalues and eigenvectors on larger volumes
($L=24,\,32,\,40$) in the vicinity of the point in the spectrum where
the localisation properties show a dramatic change, for $\beta_{\rm 
  Ising}=1.0$ and $\Lambda=1.5$. We used 20k configurations for
$L=24$, scaling down the statistics for larger volumes by keeping
approximately constant the total number of eigenvalues computed in the
relevant spectral window. In Fig.~\ref{fig:3bis} we show $I_\lambda$
as obtained from these simulations: one can easily appreciate the
tendency of the cross-over from Poisson to Wigner-Dyson statistics to
become steeper as the volume is increased. 

\begin{table}[t]
  \centering
\begin{tabular}[t]{c|c|c}
 \backslashbox[0pt][lr]{$\beta_{\rm Ising}$}{$\Lambda$}    & 1.5  & 2.0  \\
\hline 
0.65 & 1.1153(5)  & 1.4870(7) \\
0.70 & 1.1903(3)  & 1.5871(4) \\
1.00 & 1.3386(1)  & 1.7849(2)
\end{tabular}
  \caption{Value of $\Lambda\f{1+\la s_x\ra}{2}$ corresponding to the
    various choices of $\beta_{\rm Ising}$ and $\Lambda$ employed in
    this paper.}
  \label{tab:1}
\end{table}

Although the data are not of sufficiently good quality to allow a
proper finite-size-scaling analysis, it is still possible to make a
rough qualitative attempt at showing compatibility of the critical
properties with those of the three-dimensional orthogonal Anderson
model. According to the one-parameter scaling hypothesis, data from 
different volumes should collapse on a single curve if plotted against 
the scaling variable $(\lambda-\lambda_c)L^{\f{1}{\nu}}$, where
$\lambda_c$ is the critical point and $\nu$ the critical exponent of
the correlation length. The pretty large distance between the crossing
point of the $L=24,\,32$ data and of the $L=32,\,40$ data indicates
the presence of sizable finite-size corrections to the position of the
critical point, so we limit ourselves to the two largest volumes. In
Fig.~\ref{fig:3bis} we show the data collapse for the $L=32,\,40$
data. We determined $\lambda_c$ as the crossing point of second-order
polynomial fits to the data, and used the value $\nu=1.57$
corresponding to the three-dimensional orthogonal Anderson
model~\cite{nu_orth}. Although far from conclusive, this plot shows at 
least that agreement between the critical properties of the two models
is possible.

\section{Conclusions}
\label{sec:concl}

In this paper we have discussed a possible mechanism leading to the
localisation of low-lying modes of the Dirac operator in the
high-temperature phase of QCD. Our proposal is that the localising
``traps'' are provided by spatial fluctuations (``islands'') of the local
Polyakov lines away from the ordered value (i.e., the identity in
colour space). In this paper we have generalised the original proposal of
Ref.~\cite{Bruckmann:2011cc} to $SU(3)$, which has allowed us to
identify the relevant variables for localisation, namely the phases of
the eigenvalues of the local Polyakov lines, which govern the
``depth'' of the localising ``traps''. These phases enter the Dirac
equation in the form of effective boundary conditions for the quark
eigenmodes, and provide a three-dimensional source of disorder. The
three-dimensional nature of the ``islands'', which constitute
disconnected regions in a ``sea'' of (almost) ordered Polyakov lines,
sheds light on the compatibility of the critical properties of the
Dirac spectrum at the localisation/delocalisation transition with
those of the three-dimensional unitary Anderson model~\cite{crit}.

To further substantiate our proposal, and check the viability of the
``sea/islands'' explanation, we have reproduced the qualitative
features of the localised modes in a three-dimensional effective
model, designed to produce localisation precisely through the proposed
mechanism. This  ``Ising-Anderson'' model is built replacing the
four-dimensional (anisotropic) lattice, appropriate for
finite-temperature QCD, with a three-dimensional one, and replacing
the temporal covariant derivative in the Dirac operator with a
diagonal (on-site) noise. The dynamics of the diagonal noise is
governed by an Ising model with continuous spins in the ordered 
phase, in order to reproduce the qualitative features of the Polyakov
line configurations. Localised modes are indeed present, and respond 
to changes in the parameters (i.e., the temperature of
the Ising system and the strength of the spin-fermion coupling) 
as expected theoretically, thus reinforcing our confidence in the
validity of the ``sea/islands'' explanation in QCD.

There are several possible extensions of the present work. Besides a
thorough check of the critical properties of the model at the
localisation/delocalisation transition, to be compared with those of
the Anderson model in the appropriate symmetry class, the most
interesting issue to study is how to properly model the effect of the
spatial hopping. As a matter of fact, the diagonal disorder which we
believe is responsible for localisation in the high-temperature phase,
i.e., the effective boundary conditions, is present also in the
low-temperature phase of QCD, but in that case it is ineffective in
producing localisation. Indeed, in the low-temperature phase the
system looks ``fully four-dimensional'', and is basically insensitive
to the temporal boundary conditions. In our opinion, in order to
understand why this happens, and how the transition between the two
phases is realised, adopting the point of view of ``QCD as a
random-matrix model'' held in the present paper, it is necessary to
understand how the spatial hopping acts in the two phases, especially
concerning its capability of producing delocalised states around the
origin. This could lead to useful insights in the subject of chiral
symmetry breaking in QCD. 

\section*{Acknowledgements}
MG and TGK are supported by the Hungarian Academy of Sciences under
``Lend\"ulet'' grant No. LP2011-011. FP is supported by OTKA under the 
grant OTKA-NF-104034.

\appendix

\section{Symmetry class of the effective model}
\label{app:B}

The Hamiltonian of our toy model is of the form
\begin{equation}
  \label{eq:1}
  H = \gamma_4 A + i\vec \gamma \cdot \vec\de\,,
\end{equation}
with $A_{\vec x \vec y}=a_{\vec x}\delta_{\vec x \vec y}$, $a_{\vec
  x}\in\mathbb{R}$, and $(\de_j)_{\vec x \vec y} =
\f{1}{2}(\delta_{\vec x+\hat \jmath,\vec y}-\delta_{\vec x-\hat
  \jmath,\vec y})$. In this Appendix Latin indices run from 1 to 3,
Greek indices from 1 to 4. We want to find an antiunitary
``time-reversal'' symmetry of this system. We denote with ${\cal T}$
the corresponding operator, which we parameterise as ${\cal T}=K\Gamma
U$, with $K$ the usual complex conjugation and $U$ unitary to be
determined, and $\Gamma=\gamma_5\gamma_4\gamma_2$ such that
$\Gamma^\dag \gamma_\mu^*\Gamma=\gamma_\mu$, with $\Gamma^*=\Gamma$,
$\Gamma^\dag\Gamma=\mathbf{1}$ and $\Gamma^2=-\mathbf{1}$. We have to
impose 
\begin{equation}
  \label{eq:2}
{\cal T}^\dag H {\cal T} =  U^\dag \Gamma H^* \Gamma U
= U^\dag(\gamma_4 A - i\vec \gamma \cdot \vec\de)U =\gamma_4
A + i\vec \gamma \cdot \vec\de\,. 
\end{equation}
Since the last relation has to hold for any choice of $a_{\vec x}$, we see that
\begin{equation}
  \label{eq:3}
  U^\dag \gamma_4 A U = \gamma_4 A\,, \qquad  U^\dag\vec \gamma \cdot
  \vec\de U = - \vec \gamma \cdot \vec\de\,, 
\end{equation}
by setting $A=0$, and also
\begin{equation}
  \label{eq:3bis}
  U^\dag \gamma_4  U = \gamma_4 \,,
\end{equation}
by setting $a_{\vec x}=a$. $U$ must be of the general form
\begin{equation}
  \label{eq:3ter}
  U_{\vec x \vec y} = U^0_{\vec x \vec y} + U^5_{\vec x \vec
    y}\,\gamma_5 + \sum_\mu U^\mu_{\vec x \vec y}\gamma_\mu + \sum_\mu
  \tilde U^\mu_{\vec
    x \vec y}\,
  i\gamma_5\gamma_\mu + \sum_{\mu,\nu} V_{\vec x \vec y}^{\mu\nu}\sigma_{\mu\nu}\,,
\end{equation}
where $\sigma_{\mu\nu}=\f{1}{2i}[\gamma_\mu,\gamma_\nu]$. 
Imposing Eq.~\eqref{eq:3bis} restricts the form of $U$ to
\begin{equation}
  \label{eq:3quater}
  U_{\vec x \vec y} = U^0_{\vec x \vec y} +  U^4_{\vec x \vec
    y}\,\gamma_4 + \sum_j\tilde U^j_{\vec x \vec y}\,
  i\gamma_5\gamma_j + \sum_{i,j} V^{ij}_{\vec x \vec y}\,\sigma_{ij}\,.
\end{equation}
Plugging this into the first equation in Eq.~\eqref{eq:3} we obtain 
\begin{equation}
  \label{eq:3quinquies}
  (a_{\vec x}-a_{\vec y})U_{\vec x \vec y} = 0 \,,
\end{equation}
which implies
\begin{equation}
  \label{eq:3sexties}
 U_{\vec x \vec y}={\cal U}_{\vec x}\,\delta_{\vec x \vec y}\,,\qquad
 {\cal U}_{\vec x} = u^0_{\vec x} + 
 u^4_{\vec x}\gamma_4 + \sum_j \tilde u^j_{\vec x} \,i\gamma_5\gamma_j +
\sum_{i,j} v^{ij}_{\vec x}\sigma_{ij}\,. 
\end{equation}
Using now the second equation in Eq.~\eqref{eq:3} we obtain, for the
non-vanishing entries with $\vec y=\vec x\pm \hat \jmath$,
\begin{equation}
  \label{eq:4}
  {\cal U}_{\vec x\pm \hat \jmath} = -\gamma_j\, {\cal U}_{\vec x}\, \gamma_j
  ~~~\text{(no sum over $j$)}\,.
\end{equation}
This imposes the following equations,
\begin{equation}
  \label{eq:5sc}
  \begin{aligned}
     u^0_{\vec x+\hat \jmath} &= -u^0_{\vec x}\,, &&&
     \tilde u^k_{\vec x+\hat \jmath} &= -\tilde u^k_{\vec x}~~\forall k\neq j\,,&&&
     v^{kl}_{\vec x+\hat \jmath} &= -v^{kl}_{\vec x}~~\forall k,l\neq j\,, \\ 
     u^4_{\vec x+\hat \jmath} &=  u^4_{\vec x}\,, &&&
     \tilde u^j_{\vec x+\hat \jmath} &= \tilde u^j_{\vec x}\,,&&&
     v^{jk}_{\vec x+\hat \jmath} &= v^{jk}_{\vec x}\,,
  \end{aligned}
\end{equation}
which are solved by
\begin{equation}
  \label{eq:5bis}
  \begin{aligned}
     u^0_{\vec x} &= (-1)^{\sum_k x_k} u^0\,, &&&
     \tilde u^i_{\vec x} &= (-1)^{\left(\sum_k x_k\right) - x_i}\tilde u^i\,,  \\
     u^4_{\vec x} &=  u^4\,, &&&
     v^{ij}_{\vec x} &= (-1)^{\left(\sum_k x_k\right) - x_i-x_j}v^{ij}\,.
  \end{aligned}
\end{equation}
However, since $U_{\vec x \vec y}$ must respect the periodic boundary conditions,
for lattices of odd linear size the only possibility is ${\cal
  U}_{\vec x} = e^{i\phi} \gamma_4$. Computing the square of the
time-reversal operator,  
\begin{equation}
  \label{eq:6sc}
{\cal T}^2=K\Gamma U  K\Gamma U   = -\Gamma^\dag U^* \Gamma U
= - (\gamma_4)^2 = -\mathbf{1}\,,
\end{equation}
so for odd lattices the symmetry class is the {\it symplectic}
one. For even lattices, it is possible to choose ${\cal U}_{\vec x} =
e^{i\phi}(-1)^{\sum_k x_k - x_j} i\gamma_5\gamma_j$ for any $j$, which
yields
\begin{equation}
  \label{eq:6bis}
{\cal T}^2= - (\gamma_5\gamma_j)^2 = \mathbf{1}\,,
\end{equation}
so that the symmetry class is the {\it orthogonal} one (the existence
of at least one antiunitary symmetry operator with square one is
sufficient to see this). Even more simply, after spin diagonalisation
the Hamiltonian reads 
\begin{equation}
  \label{eq:1_spindiag}
  {\cal H}_{\vec x \vec y} = \eta_4(\vec x) A_{\vec x \vec y} + i\vec
  \eta(\vec x) \cdot \vec\de_{\vec x \vec y}\,, 
\end{equation}
and one can easily identify the time-reversal operator ${\cal
  T}_{\vec x \vec y}=\eta_4(\vec x) \delta_{\vec x \vec y} K$, which has ${\cal
  T}^2=\mathbf{1}$.


\newpage

\begin{thebibliography}{99}

\bibitem{BC} T.~Banks and A.~Casher, Nucl. Phys. B 
\nyp{169}{103}{1980}.

\bibitem{VWrev}
  J.~J.~M.~Verbaarschot and T.~Wettig,
  Ann.\ Rev.\ Nucl.\ Part.\ Sci.\  
\nyp{50}{343}{2000} 
  [hep-ph/0003017].

\bibitem{deF}
  P.~de Forcrand,
  AIP Conf.\ Proc.\  
\nyp{892}{29}{2007} 
  [hep-lat/0611034].

\bibitem{KP2} T.~G.~Kov\'acs and
    F.~Pittler, Phys.\ Rev.\ D 
\nyp{86}{114515}{2012}  
  [arXiv:1208.3475 [hep-lat]].

\bibitem{feri} M.~Giordano, T.~G.~Kov\'acs and
    F.~Pittler,  PoS  LATTICE 
\nyp{2013}{212}{2013}
     [arXiv:1311.1770    [hep-lat]].  

\bibitem{GGO} 
  A.~M.~Garc\'ia-Garc\'ia and J.~C.~Osborn,
  Nucl.\ Phys.\ A 
\nyp{770}{141}{2006} 
  [hep-lat/0512025].

\bibitem{GGO2}
  A.~M.~Garc\'ia-Garc\'ia and J.~C.~Osborn,
  Phys.\ Rev.\  D 
\nyp{75}{034503}{2007}
  [hep-lat/0611019].

\bibitem{KGT} T.~G.~Kov\'acs,
  Phys.\ Rev.\ Lett.\ 
\nyp{104}{031601}{2010}   
[arXiv:0906.5373 [hep-lat]]. 


\bibitem{KP} T.~G.~Kov\'acs and
    F.~Pittler, Phys.\ Rev.\ Lett.\ 
\nyp{105}{192001}{2010} 
  [arXiv:1006.1205 [hep-lat]].

\bibitem{Aoki:2005vt} 
  Y.~Aoki, Z.~Fodor, S.~D.~Katz and K.~K.~Szab\'o,
  \JHEP \nyp{01}{089}{2006} 
  [hep-lat/0510084].

\bibitem{Borsanyi:2010cj} 
  S.~Bors\'anyi, G.~Endr\H odi, Z.~Fodor, A.~Jakov\'ac, S.~D.~Katz, S.~Krieg,
  C.~Ratti and K.~K.~Szab\'o,
  \JHEP \nyp{11}{077}{2010} 
  [arXiv:1007.2580 [hep-lat]].


\bibitem{Cossu:2014aua}
  G.~Cossu {\it et al.}  [JLQCD Collaboration],
  arXiv:1412.5703 [hep-lat].

\bibitem{GKKP} M.~Giordano, T.~G.~Kov\'acs, S.~D.~Katz and F.~Pittler, 
  arXiv:1410.8392 [hep-lat].

\bibitem{unimproved}
P.~de Forcrand and O.~Philipsen, \JHEP 
\nyp{11}{012}{2008}
[arXiv:0808.1096 [hep-lat]].

\bibitem{crit} M.~Giordano, T.~G.~Kov\'acs and F.~Pittler, 
    Phys.\ Rev.\ Lett.\ 
\nyp{112}{102002}{2014} 
  [arXiv:1312.1179 [hep-lat]].

\bibitem{Anderson58}
  P.~W.~Anderson, Phys.\ Rev.\ 
\nyp{109}{1492}{1958}.

\bibitem{LR}
  P.~A.~Lee and T.~V.~Ramakrishnan,
  Rev.\ Mod.\ Phys.\  
\nyp{57}{287}{1985}.

\bibitem{EM}
  F.~Evers and A.~D.~Mirlin,
  Rev.\ Mod.\ Phys.\  
\nyp{80}{1355}{2008}
[arXiv:0707.4378 [cond-mat.mes-hall]]. 

\bibitem{Mehta}
  M.~Mehta, {\it Random Matrices}
(Academic Press, San Diego, 1991).

\bibitem{nu_unitary}
K.~Slevin and T.~Ohtsuki, Phys.\ Rev.\ Lett.\ 
\nyp{78}{4083}{1997}
[cond-mat/9704192 [cond-mat.dis-nn]].

\bibitem{offdiag} E.~N.~Economou, P.~D.~Antoniou, Solid State Commun.\
\nyp{21}{285}{1977}.

\bibitem{offdiag2}
D.~Weaire and V.~Srivastava, Solid State Commun.\
\nyp{23}{863}{1977}.

\bibitem{Bruckmann:2011cc} 
  F.~Bruckmann, T.~G.~Kov\'acs and S.~Schierenberg,
  Phys.\ Rev.\ D 
\nyp{84}{034505}{2011} 
 [arXiv:1105.5336 [hep-lat]].

\bibitem{RW}
  A.~Roberge and N.~Weiss,
  Nucl.\ Phys.\ B 
\nyp{275}{734}{1986}.

\bibitem{immu}
  P.~de Forcrand and O.~Philipsen,
  Nucl.\ Phys.\ B 
\nyp{642}{290}{2002} 
  [hep-lat/0205016].

\bibitem{immu2}
  M.~D'Elia and M.~P.~Lombardo,
  Phys.\ Rev.\ D 
\nyp{67}{014505}{2003}  
  [hep-lat/0209146].

\bibitem{GGC}   A.~M.~Garc\'ia-Garc\'ia and E.~Cuevas,
Phys.\ Rev.\ B 
\nyp{74}{113101}{2006}
[cond-mat/0602331 [cond-mat.dis-nn]].

\bibitem{Yaffe:1982qf}
  L.~G.~Yaffe and B.~Svetitsky,
  Phys.\ Rev.\ D 
\nyp{26}{963}{1982}.

\bibitem{DeGrand:1983fk}
  T.~A.~DeGrand and C.~E.~DeTar,
  Nucl.\ Phys.\ B 
\nyp{225}{590}{1983}.

\bibitem{nu_orth}
K.~Slevin and T.~Ohtsuki, Phys.\ Rev.\ Lett.\ 
\nyp{82}{382}{1999} 
[cond-mat/9812065 [cond-mat.dis-nn]].


\end{thebibliography}
\end{document}